\documentclass[twocolumn,pra,aps,floatfix,showpacs]{revtex4-2}
\usepackage{epsfig,graphicx,tabularx,epstopdf}

\begin{document}
\title{Rotation-driven transition  into coexistent  Josephson modes in an atomtronic dc-SQUID}
\author{D. M. Jezek and H. M. Cataldo}
\affiliation{IFIBA-CONICET
\\ and \\
Departamento de F\'{\i}sica, FCEN-UBA
Pabell\'on 1, Ciudad Universitaria, 1428 Buenos Aires, Argentina}

\begin{abstract}
By means of a two-mode model, we show that transitions to different arrays of coexistent regimes in the phase space can be attained by rotating a double-well system, which consists of a toroidal condensate with two diametrically placed barriers. Such a configuration corresponds to the atomtronic counterpart of the well-known direct-current superconducting quantum interference device. Due to the phase gradient experimented by the on-site localized functions when the system is subject to rotation, a phase difference appears on each junction in order to satisfy the quantization of the velocity field around the torus. We demonstrate that such a phase can produce a significant change on the relative values of different types of hopping parameters. In particular, we show that within a determined rotation frequency interval, a hopping parameter, usually disregarded in nonrotating systems, turns out to rule the dynamics. At the limits of such a frequency interval, bifurcations of the stationary points occur, which substantially change the phase space portrait that describes the orbits of the macroscopic canonical conjugate variables.  We analyze the emerging dynamics that  combines the $0$ and $\pi$ Josephson modes, and evaluate the small-oscillation time-periods of such orbits at the frequency range where each mode survives.  All the findings  predicted by the model are  confirmed by Gross-Pitaevskii simulations.
\end{abstract}
%
%
\maketitle
\section{Introduction}
Much interest has been devoted in recent years to Bose-Einstein condensates confined by toroidal traps radially crossed by a number of rotating barriers, as such configurations present the
clear potential of becoming central building blocks for the future
atomtronic devices \cite{amico2021state,*acircuits}. 
In case of a single rotating barrier, which yields the cold atom analogue of the celebrated rf-SQUID
(radio frequency-Superconducting Quantum Interference Device)--- a superconducting ring interrupted by a Josephson junction
\cite{braginski,*fagaly},
well-defined phase slips between quantized persistent currents have been observed \cite{wright}, along with a quantized
hysteresis behavior \cite{eckel}. The case of two diametrically disposed barriers corresponds to the cold atom version of
the dc-SQUID \cite{braginski,*fagaly}, perhaps the most sensitive detector for magnetic flux available today.
We may denote such atomtronic counterparts of SQUIDs as AQUIDs, for Atomtronic Quantum Interference Devices
\cite{amico2021state,*acircuits}. In a SQUID, a current flow is established by changing the magnetic flux through the loop,
whereas the same effect in an AQUID is obtained as a consequence of the barrier rotation,
or, equivalently, by imparting a geometric phase directly to the atoms via suitably designed laser fields \cite{rmp}. 
This makes rotation sensing possible,
as already demonstrated for neutral atoms of superfluid helium \cite{satop} and very recently for ultracold atoms
\cite{ryu20}, where the quantum interference of currents in a dc-AQUID
was observed. Such efforts, along with another recent proposals to achieve
rotation sensing by atomic persistent currents \cite{pele,nico,kumar},
 may be considered as part of the rapidly growing branch of research known as `quantum sensing' \cite{degen}. 

 Previously, Josephson effects \cite{boshier13} and resistive flows \cite{jen14} had also been observed in dc-AQUIDs. On the other hand,
a variant of rf-AQUID constructed upon a ring-shaped optical lattice interrupted by a weak link has been 
proved to yield an effective two-level
quantum dynamics able to reproduce a qubit \cite{rf-aquid}. Later, it was shown that the
same lattice confinement, but interrupted by three
weak links, also reproduces an effective qubit dynamics, 
but in a considerably enlarged parameter space \cite{3-wl}.
Such efforts are opening the way towards an experimental realization of
a cold atom analogue of the superconducting flux qubit \cite{amico14}. 

From the theoretical point of view, the modelization of AQUIDs
has mostly been addressed by one-dimensional (1D), i.e.
tight-waveguide, descriptions for which the Lieb-Liniger model generalized to host barrier potentials
can be applied \cite{caza,polo}. Different methods depending on the interaction strength have been
utilized to study these systems. In the limit of weak interactions, the relevant physics of the system
can be captured within the mean-field approximation by the Gross-Pitaevskii (GP) equation 
\cite{polo,polo1,rf-aquid},
whereas in the hard-core limit of infinite
repulsion, the Tonks-Girardeau Bose-Fermi mapping often provides analytic solutions \cite{caza,polo,rf-aquid}.
For interaction strengths outside the above limits, one may resort to various computational techniques, such as
several kinds of Monte Carlo approaches \cite{caza}, and in case of ring lattices,
exact diagonalization schemes for small systems and
the density-matrix renormalization group for larger system sizes \cite{caza,rf-aquid}. Another useful approach
comes from the quantum phase model, which assumes a system dynamics characterized
by the phase differences across each junction, neglecting the site number fluctuations
 \cite{fazio,rf-aquid}.

In the case of nonrotating double-well systems, a two-mode (TM) model  was developed 
in Refs. \cite{smerzi97,ragh99}
 and latterly improved by considering terms that gave rise to novel parameters \cite{anan06}.  
We will focus in this paper particularly on one of such parameters, which has shown to provide almost vanishing effects  in the nonrotating context.
 Given that a discrepancy still persisted between the model results and the Gross-Pitaevskii (GP) simulations, an effective TM model was developed in recent years \cite{cap13}.  Such a model arises from introducing
  in the derivation of the equations of motion the 
on-site interaction energy dependence on the population
imbalance. 
When doing so, one obtains the same type of equations, but with an effective interaction energy parameter
instead of the standard on-site interaction one.
In the Thomas-Fermi approximation, it has been shown that the effective interaction energy parameter 
becomes reduced with respect to the standard one by a factor of 7/10, 3/4, or 5/6, depending on
the dimensionality of the system. 
Such an effective TM model has shown  to accurately describe the exact dynamics
in double-well \cite{cap13,nigro17}, asymmetric two-well \cite{cat14,cat20} and, in the multimode version,
multiple-well \cite{je13,nigro18} condensates. 

A  multimode model,  which includes all the type of corrections introduced in  the TM model, 
was recently proposed for describing the dynamics of cold atoms confined by a rotating  ring-shaped optical lattice forming a weakly linked array of condensates \cite{nigro20}.
The on-site localized functions were obtained by means of a change of basis from that formed by
the stationary order parameters with different winding numbers. Due to  the rotation, the localized functions acquire a phase gradient along the bulk,
which causes the formation of a phase difference across the junctions of
neighboring localized functions. It has been shown that such a phase difference
turns out to determine the argument of the complex hopping parameters.  

In this paper, we will apply the above multimode
model  restricted to a double-well system in the form of a rotating TM version.
Even though all the model parameters will be shown to be real numbers in this case, we will see 
that the phase gradient imprinted on the on-site localized functions, similarly to the multi-well case,
plays a crucial role in determining the behavior of the hopping amplitudes as functions of the frequency. 
In particular, we will see that there exist frequencies where the standard hopping amplitudes vanish, and a usually disregarded parameter for nonrotating systems turns out to define the dynamics. 
Such a parameter, which in the nonrotating case is defined as proportional to the integral of the product of 
the densities of both localized functions, 
 splits for the rotating condensate into two different model parameters, one of which of a hopping-like nature, playing an essential role,  while the other 
of interaction-like character, yielding negligible corrections.

One goal of this work consists in showing that in correspondence with the increment of the role played by
such a hopping parameter, the system exhibits the coexistence of two Josephson oscillation modes. In particular, we show that at the extremes of the frequency interval where both modes coexist, bifurcations of the stationary points take place. Thus, we analyze the way in which rotation affects the values of the model parameters, and further elucidate how such parameters modify the phase portrait and  time periods.
Such essential
insights provided by the TM model
are tested against GP
simulations, which confirm us that we are dealing with a simple and quite accurate theoretical tool
accounting for the response to
rotation of a dc-AQUID. 


This paper is organized as follows. In Sec. \ref{sec2} we describe the system and details of the GP simulations we
have performed. Section \ref{sec3} deals with the rotating TM model, where
we analyze the phase gradient on the on-site localized functions and derive the equations
of motion in Sec.~\ref{sec3aa}, while the dependence of the model parameters on the rotation frequency is discussed in Sec. \ref{sec3a}.
The phase portrait is described in Sec. \ref{sec4}, where
we study the distribution of stationary points with their possible bifurcations in Sec. \ref{sec4a},
whereas the corresponding different types of phase portraits, particularly those presenting a coexistence of Josephson modes, are analyzed in Sec.~\ref{sec4b}. Section \ref{sec5} is devoted to examine the dynamics of orbits on the different
regimes. We pay special attention to the time periods of both Josephson modes, deriving
an analytical expression in the small-oscillation (SO) approximation. All the evolutions are compared to
time-dependent GP simulations. Some concluding remarks are gathered in Sec.~\ref{sec6}. 
Finally, useful alternative calculations of the hopping parameters are summarized in the Appendix. 

\section{Theoretical framework}\label{sec2}
We describe in what follows the condensate we have considered in our study, which was experimentally
realized as AQUIDs in Refs. \cite{boshier13,ryu20}.
The trapping potential can be written as the sum of a term depending
on $x$ and $y$ and a term that is harmonic in the tightly bound direction $z$,
\begin{equation}
V_{\text{trap}}=V(x,y)+\lambda^2z^2/2
\label{poten}
\end{equation}
with
\begin{equation}
V(x,y)=V_0\left[1-\left(\frac{r^2}{r_0^2}\right)\,\exp\left(1-\frac{r^2}{r_0^2}\right)\right]
+V_b\exp(-y^2/\lambda_b^2).
\label{poten1}
\end{equation}
The above potential consists of a superposition of
a toroidal term modeled through a Laguerre-Gauss optical potential
\cite{lag},
where $V_0$ corresponds to the potential depth and $r_0$ the radial position
of its minimum ($r^2=x^2+y^2$),
and a Gaussian barrier along the $x$ axis of height $V_b$ and width $\lambda_b$ that splits
the torus into two halves, the one on the upper half-plane labeled as the `$u$' site and the
other located on the lower-half plane, named the `$l$' site.
The following system parameters were considered \cite{boshier13},
  $V_0$ = 70 nK, $r_0$ = 4 $\mu$m,
$ V_b$ =  41 nK, $ \lambda_b = 1\, \mu$m and $N$=3000 atoms of $^{87}$Rb. 
We have also assumed a relative high value of $\lambda$,
$18.8$ nK$^{1/2}\mu$m$^{-1}$, which yields
a quasi-bidimensional condensate.
Thus, the condensate order parameter is written as the product
of a Gaussian wave function along the $z$ coordinate,
and a two-dimensional (2D) wave function $\psi({\bf r},t)$ normalized to one,
for which the corresponding GP equation in a rotating frame at the angular velocity $\Omega\hat{\bf z}$ 
reads \cite{castin},
\begin{equation}
[\hat{H}_0+gN|\psi({\bf r},t)|^2-\Omega \hat{L}_z]\psi({\bf r},t)=i\hbar\frac{\partial \psi({\bf r},t)}{\partial t},
\label{gp}
\end{equation}
where $\hat{H}_0=-\frac{\hbar^2}{2m}\nabla^2+V$ is the noninteracting Hamiltonian,
$\hat{L}_z$ denotes the $z$ component of the angular momentum operator and $g$ corresponds to
the effective $2D$ coupling constant between the atoms \cite{castin}.
Such a GP equation has been numerically solved using the 
split-step Crank-Nicolson algorithm for imaginary- and real-time propagation
on a 2D spatial grid of 257$\times$257 points \cite{kumar19}.

\section{Rotating two-mode model}\label{sec3}
\subsection{On-site localized functions and equations of motion}\label{sec3aa}

In our TM model, the condensate order parameter $\psi_{TM}({\bf r},t)$
is written in terms of a pair of order parameters, which,
by analogy with multiple-well systems, will be referred to as `on-site' localized functions.
Therefore,  being the barriers localized along $y=0$ (Eq. (\ref{poten1})),
we have an `upper' and a `lower' localized function, $\psi_u({\bf r})$ and  $\psi_l({\bf r})$,
respectively, from which the condensate order parameter is built as,
\begin{equation}
\psi_{TM}({\bf r},t)=b_u(t)\psi_u({\bf r})+b_l(t)\psi_l({\bf r}),
\label{psiTM}
\end{equation}
with $b_k(t)=\sqrt{N_k(t)/N}\exp[\varphi_k(t)]$ ($k=u,l$), where 
$N_k(t)$ represents the number of particles at the site $k$
and $\varphi_k(t)$ denotes a global phase that takes into account the time dependence of the phase
on that site \cite{nigro20}. Note that the spatial coordinate dependence of the condensate order parameter
stems from those of the on-site localized
functions. These are complex functions, the phases of which, respectively denoted as $\phi_u({\bf r})$ and $\phi_l({\bf r})$
for the upper and lower localized functions,
will be discussed below. In the nonrotating case,
the localized functions arise from
 the sum and difference of the lowest energy stationary order parameters
obtained from the GP equation, i.e., the ground state with winding number $n=0$ and
the antisymmetric state, which for our toroidal configuration corresponds to 
a winding number $n=1$ \cite{je13}.
We note that the stationary order parameter of maximum winding number for nonrotating  ring lattices, with an even number of sites, is characterized by presenting nodal surfaces along the barriers. The phase  is homogeneous in each site, except at the junctions where a $ \pi $  phase difference  exists. Hence, the uniform phases alternate between $0$ and $\pi$ in consecutive sites.  In our case the two-site lattice has a maximum winding number  of 1, and hence the associated order parameter has the same behavior of an antisymmetric one.

For a finite angular frequency $\Omega$, one may first calculate
the rotating stationary order parameters $\psi_0$ and $\psi_1$ of winding numbers  0 and 1, 
respectively, by imaginary-time propagating  through the GP equation (\ref{gp})
the corresponding nonrotating stationary states.
It is worthwhile noticing that the total phase difference of $\psi_1$  around the torus  is now distributed not only in the junctions but also in the bulk.
Since such states have different initial winding numbers, the imaginary-time propagation
keeps the orthogonality between them. This represents an important feature, as it determines the
conservation of the two-state orthonormal basis in the rotating configuration.
We may obtain the basis formed by the on-site localized functions as the sum and difference of the rotating stationary
order parameters $\psi_0$ and $\psi_1$. It is important to note that in order to achieve a maximum localization, such stationary states should have
their phases previously fixed to zero at the middle of a site
\cite{nigro18}, which we have chosen to be the upper one, and 
at the point  $(x=0;y=r_0)$. 
\begin{figure*}
\includegraphics{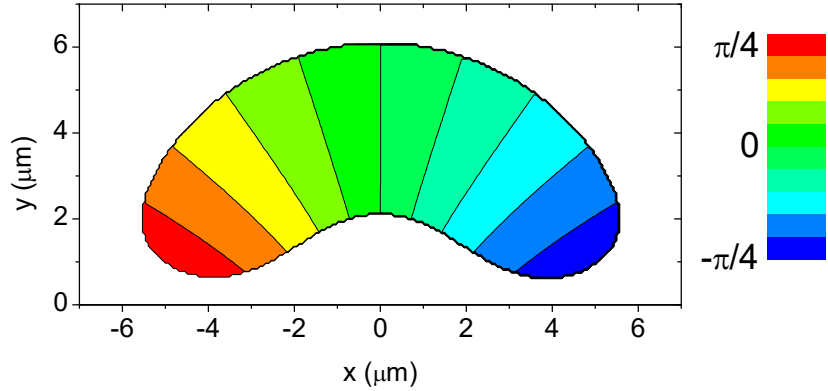}
\caption{Isodensity contour of the on-site localized function of the upper site 
$\psi_u(\bf r)$ and its corresponding phase $\phi_u(\bf r)$ for the frequency $f=
\Omega/2\pi=3.54$ Hz. The color scale corresponds to the phase of the function,
while the contour corresponds to 1/14 of the maximum value of $|\psi_u|^2$.}
\label{figu5}
\end{figure*}
Such an ansatz for obtaining the localized functions corresponds to that of
a rotating ring-shaped optical lattice described in Ref.~\cite{nigro20}, applied in this case to a 
two-well system.

Given that the bulks of the
rotating stationary states acquire a phase gradient  around the condensate, 
a phase with the same behavior is imprinted on the 
corresponding localized functions, as observed in Fig.~\ref{figu5}.
The upper localized function shown in this figure corresponds to the sum of both stationary states.
\begin{figure*}
\includegraphics{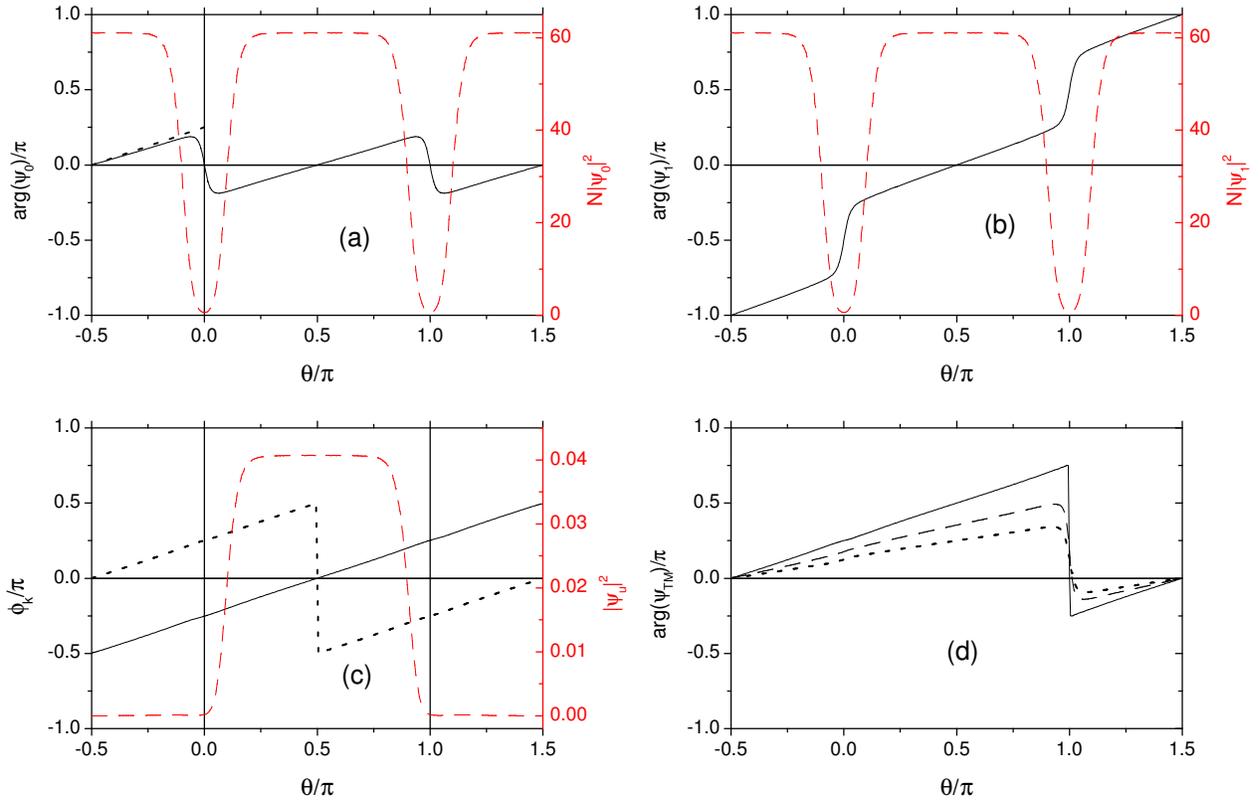}
\caption{(a): Phase (black solid line) and density (red dashed line) of the stationary state $\psi_0$ of winding number 0
for the condensate rotating at the frequency $f=\Omega/2 \pi= 3.54$ Hz.
The dotted line displays the tangent to the phase at $\theta/\pi=-0.5$, the intersection of which with the  solid  vertical line indicates half of the phase difference  across the junction. Such a vertical solid line shows the location of the right junction.
(b): Solid and dashed lines: same as panel (a) for the stationary state $\psi_1$ of winding number 1. (c): Phase $\phi_u$ of the upper (solid line) and lower $\phi_l$ (dotted line)
on-site localized functions derived from the above stationary states.
The red dashed line corresponds to the square amplitude of the upper localized function, while the solid vertical
lines show the location of the junctions.
(d): Phase of the TM order parameter (\ref{psiTM}) with the global phases 
 $\varphi_u= -\Theta$ and $\varphi_l= 0$ and equal populations  for
three rotation frequencies: 3.54 Hz (solid line), 2.5 Hz (dashed line) and 1.77 Hz (dotted line). }
\label{figu8}
\end{figure*}
In Fig. \ref{figu8} (a) and (b)  we further show the phase and density of  the stationary states $\psi_0$ and $\psi_1$  as functions of the angular coordinate $\theta$,
at $r=r_0$, for the frequency $f=\Omega/2 \pi= 3.54$ Hz.
 In panel (c)  we show the phases, $\phi_u(\theta)$ and $\phi_l(\theta)$,
 for the corresponding localized functions,  
together with the square amplitude of the upper localized function.
There it may be seen that a phase difference  $\Delta \phi_u= \pi/2$  
exists  between $\theta =0$ and $ \theta= \pi$ of such localized function. 
To qualitatively explain such a value, it is useful to
 reduce the treatment of our system to  a rotating 1D annulus of radius $r_0 = 4\,\mu$m 
with  negligible barrier widths. 
In such a case, any fluid element should move with an angular velocity $\Omega$,
and hence the phase difference between the ends of each semicircle yields the 1D prediction 
$\Delta \phi_{1 D}= \pi  \Omega r_0^2 m/ \hbar$. Then,  for this simplified model, 
the $\pi/2$ difference should be attained for a rotation frequency of 3.6 Hz. On the other hand,
in our more realistic extended system, the velocity field verifies the superfluidity condition and then acquires a more complex form \cite{nigro20}. 
Hence, we have found that the phase difference of $\pi/2$ is attained at the slightly smaller frequency of 
3.54 Hz, with respect to the 1D prediction. We will see that such a phase difference on the on-site localized functions involves  the presence  of a related  phase difference at the junctions. First, we note that it is easy to verify that  the stationary state  with winding  number $n=0$ corresponds  to  the TM order parameter (\ref{psiTM}) with  vanishing global phases and equal populations $N_u=N_l$. 
However, in spite of the vanishing global phases,
in the panel (a) of Fig. \ref{figu8} an extra negative  phase difference can be appreciated at  the junctions, which will be denoted as $ \Theta$.
In order to obtain a quantized velocity field circulation around the torus, such a phase should verify  $ \Theta = -\pi/2$, which for any frequency can be generalized to
  $ \Theta = -\Delta \phi_u$. With the purpose of better visualize the precision of the $\Theta$ value in
  Fig.~\ref{figu8} (d), we show as a solid line the phase  of the TM order parameter (\ref{psiTM})
 for global phases 
 $ \varphi_u=\pi/2= - \Theta$ and $ \varphi_l=0$ with identical populations at each site.
It may be seen that the discontinuity disappears indeed at the right junction ($\theta=0$), 
whereas a phase of $-\pi$ is present
 at the left junction, as expected. 
The same procedure was applied to two other rotation frequencies,
showing that the phase difference at the right junction cancels indeed.
For these two frequencies
we have assumed a linear dependence for $ -\Theta(f)$, with $\Theta(3.54\, {\rm Hz})=-\pi/2$, 
although,  in practice, such a slope presents a small  increment with the frequency, and hence
this approximation can only be taken locally.
The importance of presence of the phase $ \Theta$
 will become evident when analyzing the behavior of the TM model hopping parameters.
 
Finally, by inserting the order parameter (\ref{psiTM}) in the GP equation (\ref{gp}), 
we may extract after some algebra the TM equations of motion 
in terms of the particle imbalance $Z=(N_l-N_u)/N$ and
the global phase difference between both sites of the condensate $\varphi=\varphi_u-\varphi_l$,
\begin{equation}
\hbar\frac{dZ}{dt}=-2K\sqrt{1-Z^2}\sin\varphi+\varepsilon(1-Z^2)\sin2\varphi,
\label{zpun}
\end{equation}
\begin{equation}
\hbar\frac{d\varphi}{dt}=Z\left[NU_{\rm eff}+\frac{2K}{\sqrt{1-Z^2}}\cos\varphi-\varepsilon\cos 2\varphi-2\varepsilon '\right],
\label{fipun}
\end{equation}
Here the TM model parameters are defined as follows \cite{smerzi97,ragh99,anan06,cap13,nigro17},
\begin{equation}
J=-\int  d^2r\,\,\, \psi_u^*(\hat{H}_0-\Omega \hat{L}_z)\psi_l,
\label{jota}
\end{equation}
\begin{equation}
F=-g N\int  d^2r\,\,\, \psi_u^*|\psi_u|^2\psi_l,
\label{efe}
\end{equation}
where $J$ and $F$ denote the standard and interaction-driven hopping parameters, respectively, the sum of which yields
the full hopping amplitude $K$. The remaining parameters are,
\begin{equation}
U=g\int  d^2r\,\,\,|\psi_u|^4,
\label{UU}
\end{equation}
\begin{equation}
\varepsilon=g N\int  d^2r\,\,\, (\psi_u^*\psi_l)^2,
\label{eps}
\end{equation}
\begin{equation}
\varepsilon '=g N\int  d^2r\,\,\, |\psi_u|^2|\psi_l|^2.
\label{epsp}
\end{equation}
 The effective on-site interaction parameter $U_{\rm eff}$ in Eq. (\ref{fipun})
arises from considering $U$ in Eq. (\ref{UU}) as a function of the imbalance.
Such an effective parameter can be written as
$U_{\rm eff}=(1-\alpha)\, U$, where the procedure to calculate $\alpha$ for an arbitrary number of particles
is explained in Refs. \cite{je13,nigro17}. The Thomas-Fermi prediction \cite{cap13} represents a lower bound
for the $1-\alpha<1$ value.
On the other hand,
$\varepsilon$ and $\varepsilon'$ respectively denote the correlated hopping amplitude and the intersite interaction
parameter.  We have adopted the above denominations for the hopping parameters in accordance with Ref.
\cite{anan06}, where it is shown that the second-quantized version of their improved  (nonrotating) TM model has a term stemming from a nonvanishing value of $\epsilon=\epsilon'$ that  contains correlated hopping or two-particle tunneling \cite{duttar}
effects, whereas what we have called the full hopping amplitude $K$, 
only involves one-particle processes.
 It is worthwhile noticing that when the system is subject to rotation, the on-site localized functions are intrinsically complex, and hence one can identify a pure interaction parameter  
and a hopping amplitude, which are given by Eqs. (\ref{epsp}) and (\ref{eps}), respectively.
In contrast, we recall that in the nonrotating case a single parameter is obtained,
since for real localized functions such expressions coincide.
We note that in the derivation of Eqs. (\ref{zpun}) and (\ref{fipun}), we have used 
the fact that for a double-well condensate 
all the model parameters turn out to be real, as will be shown in the next Subsection.

\subsection{TM model parameters}\label{sec3a}
For a rotating ring-shaped lattice with a number of sites larger than two, the parameters $J$, $F$ and
$\varepsilon$ become complex numbers, the phases of which are related to the imprinted phase on the localized
sites \cite{nigro20}. However, when only two sites with two junctions are considered, it is easy to 
verify that such parameters are real numbers. For instance, calling $J_{ul}$ the parameter defined by
Eq. (\ref{jota}), where the subscripts correspond to the order of the localized functions in the integrand,
and taking into account that $\hat{H}_0$ and $\hat{L}_z$ are hermitian operators,
we have $J^*_{ul}=J_{lu}$. On the other hand, as a result of the symmetry of the 
system, we have that
the hopping amplitude must verify $J_{ul}=J_{lu}=J$ and hence, $J^*_{ul}=J_{ul}$. The same analysis
can be applied to the remaining parameters, $F$, $\varepsilon$ and $K$. 

\begin{figure*}
\includegraphics{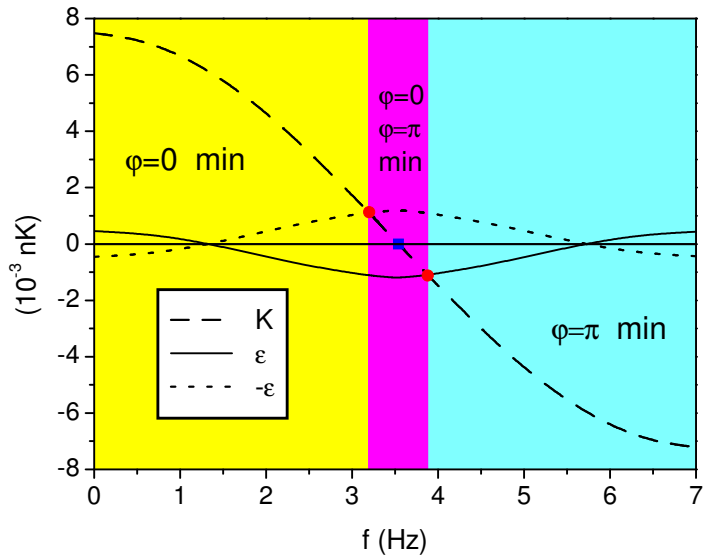}
\caption{TM parameters $K$ and $\varepsilon$ versus the rotation frequency $f=\Omega/2\pi$. 
The three frequency intervals indicated with different colors (grayscale values), 
correspond to qualitatively distinct phase space
portraits, which are characterized by the number and position of the energy minima. 
The intersection points $K=\pm\varepsilon$ (red circles) define the bifurcation frequencies $f_1=3.18$ Hz and $f_2=3.90$ Hz
which determine the limits of the central frequency interval where coexist two
energy minima at $\varphi =0$ and $\varphi =\pi$.
At the central frequency (3.54 Hz), where $K=0$ (denoted by a blue square),
such minima possess the same depth.
The remaining two frequency intervals, $ f<f_1$ and $ f>f_2$,
exhibit a single minimum, as indicated in the
drawing.}
\label{figu1}
\end{figure*}

A strong dependence
on the rotation frequency is expected for those parameters that depend
on the imprinted phase of the localized
functions. Such is the case for $K$ and $\varepsilon$, which are depicted as functions of the frequency in
Fig. \ref{figu1}. The relative values of these parameters
will become crucial at determining the nature of the stationary
points on different frequency intervals.

In order to qualitatively explain the behavior of $K$ and $\varepsilon$ as functions of the frequency,
 one can separately analyze
the origin of both terms at the right-hand side of the equation of motion (\ref{zpun}). 
First, we may consider that $\dot Z$  
has two contributions coming from the currents,   $I_L$ and $I_R$, flowing through
the left  and right  junctions, respectively. In turn, each current has,
in analogy with a single junction double-well system,
two terms, one  current  $I_i^{(1)}$
proportional  to $\sin(\varphi_i)$, 
 and the other current  $ I_i^{(2)}$  proportional to $\sin(2\varphi_i)$, where $\varphi_i$ denotes the phase difference at the right ($i=R$) and left ($i=L$)  junctions, respectively.
 In view of  the fact  that, as described in the previous subsection,  an additional phase difference
$\Theta$  appears  at each junction, we obtain that the phase differences at each junction are  $ \varphi_R =\varphi  + \Theta$  and  $ \varphi_L =\varphi  -\Theta$.
Hence, denoting as $\dot{Z}_1$ the time derivative of the imbalance due to the sum of  the currents  $I_R^{(1)}$ and  $I_L^{(1)}$ we have,
\begin{equation}
\dot Z_1 \propto \sin(\varphi + \Theta) + \sin(\varphi  - \Theta) = 2  \cos(\Theta) \sin(\varphi).
\end{equation}
Since $\dot Z_1$ stems from the term of Eq. (\ref{zpun}) proportional to $K$, 
we may conclude that the parameter $K$ should be modulated by $\cos(\Theta)$,
and the same modulation should apply for $J$ and $F$.

Analogously,  we may obtain the time derivative of the imbalance  $\dot{ Z}_2$
 due to  the sum of  the currents $I_R^{(2)}$ and  $I_L^{(2)}$ as,
\begin{equation}
\dot Z_2 \propto \sin(2(\varphi + \Theta)) + \sin(2(\varphi - \Theta)) =
2  \cos(2\Theta) \sin(2\varphi),
\label{2theta}
\end{equation}
which added to $\dot Z_1$ yields the total imbalance derivative 
 $ \dot{Z}= 
\dot{Z}_1+ \dot{Z}_2 $. Thus, from Eqs.~(\ref{2theta}) and (\ref{zpun}), we may conclude that the parameter $\varepsilon$ 
should be modulated by $\cos(2\Theta)$.

One may further use the 1D approximation  $-\Theta\simeq\Delta\phi_{1 D}= \pi  \Omega r_0^2 m/ \hbar$ and define $f_0= \hbar/(m 2 \pi r_0^2)$,
from which we have 
$K \simeq B_K(f)  \cos( f \pi/ f_0)$ and $\varepsilon \simeq  B_\varepsilon(f)   \cos( 2 f \pi/ f_0)$,
where, as seen from Fig. \ref{figu1},  the amplitudes $B_i(f)$ are positive and do not vary substantially as functions of $f$. 
For our system, the linear relationship between $\Theta$ and 
$f$ is less rigorous since its modulus  has a slightly increasing slope, but, as seen from the graph, such an approximation certainly captures the main behavior.

In nonrotating systems, the parameters $F$ and $\varepsilon$ were first 
introduced by Ananikian and Bergeman \cite{anan06} in order to improve the TM model,
however, as discussed by the authors, the latter in general verifies 
$\varepsilon<< K$. 
Hence, in most cases it is not taken into account, or at least it is only used to introduce  small corrections \cite{nigro17}. 
In this work the nonrotating value of $\varepsilon$ turns out to be about an order of magnitude smaller than $K$.

The above modulation of $K$ implies that it should vanish at   $ \Theta = -\pi/2$,
 which is attained for $ f=3.54$ Hz. On the other hand, at the same frequency,  the hopping parameter
$\varepsilon$ should acquire its maximum absolute value. Then, we infer that around such $\Theta$ value, $\varepsilon $ turns out to be the leading hopping parameter of the model. As a consequence of this effect, in the next Section it will be shown that two energy minima coexist in the phase portrait in the vicinity of such a frequency.
It is worthwhile noticing that the same type of phenomenon also occurs at,  for instance, $ \Theta = - 3/2 \pi$ for a rotation frequency of 10.54 Hz.

On the other hand, we have found that
the parameters $U$ and $\alpha$ remain almost constant when varying the 
rotation frequency, 
as also observed in a previous work \cite{nigro20}. 
In particular, we have utilized the values
$U=1.186\times 10^{-2}$ nK  and $1-\alpha= 0.814$ 
for the whole frequency interval.
Given that we have assumed a rather low number of particles, we could not make
use of Thomas-Fermi estimates,
and hence the value of $\alpha$ 
was numerically obtained,
as described in Refs. \cite{je13,nigro17}. We note that such a value lies in between the 1D
and 2D Thomas-Fermi predictions.

Finally, we note that in the Appendix we discuss about alternative formulas for obtaining the TM parameters $K$ and $\varepsilon$, in addition to
the above expressions (\ref{jota}), (\ref{efe}) and (\ref{eps}). 
There we show that in the case of $\varepsilon$,
 such alternative calculations can be used to provide results with a better accuracy.   

\section{Phase portrait}\label{sec4}
The equations of motion (\ref{zpun})-(\ref{fipun}) can be written in Hamiltonian form as,
\begin{equation}
\dot{Z}=-\partial H/\partial\varphi\,\,\,\,\, {\rm and}\,\,\,\,\,\dot{\varphi}=\partial H/\partial Z,
\label{ham}
\end{equation}
where the classical Hamiltonian $H$ 
depends on the canonical variables $Z$ and $\varphi$ as,
\begin{eqnarray}
H(Z,\varphi)&=&\left(\frac{NU_{\rm eff}}{2\hbar}-\frac{\varepsilon'}{\hbar}\right)Z^2-
2\frac{K}{\hbar}\sqrt{1-Z^2}\cos\varphi\nonumber\\
&+&
\frac{\varepsilon}{2\hbar}(1-Z^2)\cos2\varphi.
\label{hamilt}
\end{eqnarray}

\subsection{Stationary points}\label{sec4a}
According to (\ref{ham}),
the stationary points correspond to the
condition of vanishing partial derivatives of the Hamiltonian with respect to the canonical variables.
For a nonrotating condensate \cite{smerzi97,ragh99,anan06,cap13,nigro17}, 
given that $\varepsilon<K$, the stationary
points consist of an energy minimum at ($Z=0,\varphi=0$), a saddle at ($Z=0,\varphi=\pi$),
and two maxima at ($Z\simeq\pm 1,\varphi=\pi$). However, these maxima 
do not possess stationary state counterparts in
the GP equation, and hence they present no physical interest. New features appear when the condensate is 
subject to rotation, as a consequence of the fact that the value of $K$ decreases and the absolute value
of $\varepsilon$ can become larger than $K$ (see Fig. \ref{figu1}).
We will see that stationary point bifurcations occur at certain frequencies $f_1$ and $f_2$, and in
between them two energy minima coexist.
In particular, the first bifurcation frequency $f_1=3.18$ Hz is attained  
where $K=-\varepsilon$, a condition which makes the second derivative $\partial^2H/\partial\varphi^2$ to vanish at ($Z=0, \varphi=\pi$). This gives rise to a bifurcation along the $Z\equiv 0$ axis
of the kind $s\rightarrow 2s+m$, where
$s$ denotes a saddle and $m$ a minimum. Then,
at increasing frequencies, such a relative minimum stays at $\varphi=\pi$, while the saddles,
located at $\varphi_s =\pm\cos^{-1}(K/\varepsilon)$, move
in opposite directions along the $Z\equiv 0$ axis towards the point $\varphi=0$. 
Particularly, when the saddles reach
$\varphi=\pm\pi/2$ at the frequency of 3.54 Hz, the minima at $\varphi=0$ and $\varphi=\pi$ have the same energy, 
whereas for larger frequencies the minimum at
$\varphi=\pi$ acquires a lower energy than the minimum at $\varphi=0$. 
Finally, at the second bifurcation frequency for which
$\varepsilon=K$, $f_2=3.90$ Hz,  the second derivative $\partial^2H/\partial\varphi^2$
again vanishes, provoking a bifurcation of the type $2s+m\rightarrow m$. In other words,
the saddles collapse at the origin, along with the
relative minimum, yielding a single saddle at $\varphi=0$ for larger frequencies,
in addition to the minimum at $\varphi=\pi$.

\begin{figure*}
\includegraphics{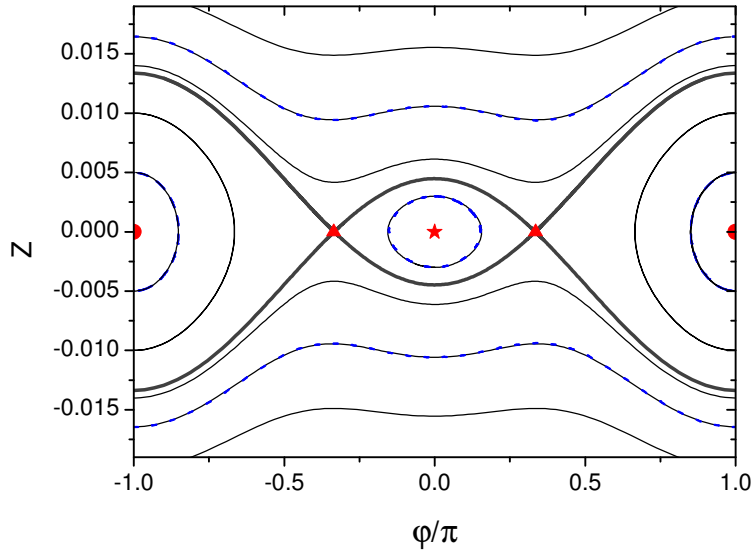}
\caption{Phase space portrait $Z$ versus $\varphi$ arising from the Hamiltonian (\ref{hamilt})
for the rotation frequency $f=3.72$ Hz. Each orbit is represented by a thin solid line, except
for the separatrix that is depicted as a thick solid line.
The circles and the star respectively represent the location of absolute and relative minima, while the triangles
correspond to saddles. A very good agreement with a few selected orbits corresponding
to GP simulations is observed (blue dashed lines).}
\label{figu2}
\end{figure*}

We display in Fig.~\ref{figu2} the phase portrait for the
rotation frequency $f=3.72$ Hz, which belongs to the central interval of Fig.~\ref{figu1}, located in
between the bifurcation frequencies $f_1$ and $f_2$. We may observe that it presents an
absolute and a relative minimum at $\varphi=\pi$ and $\varphi=0$, respectively, and a couple of saddles
at $\varphi_s =\pm\cos^{-1}(K/\varepsilon)\simeq\pm\pi/3$ ($\varepsilon\simeq 2\, K$ for this frequency). 

\subsection{Dynamical regimes on the different types of phase portraits}\label{sec4b}

\begin{figure*}
\includegraphics{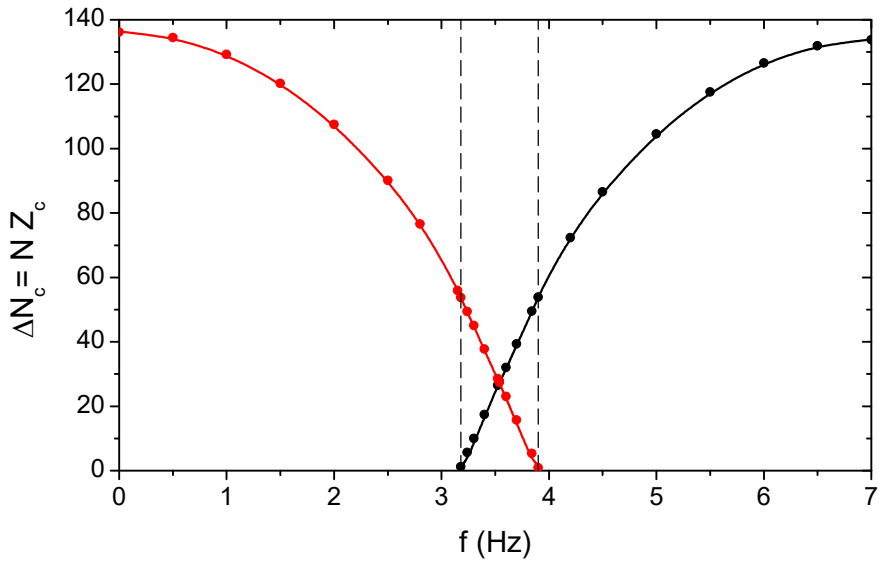}
\caption{Critical particle imbalance versus the rotation frequency. The vertical dashed lines 
indicate the bifurcation frequencies, which separate the
three frequency intervals of Fig. \ref{figu1}. The circles denote the values of GP simulation results, while the full
lines correspond to the theoretical predictions (\ref{zc1}) and (\ref{zcpm}). The results belonging to 
0-modes ($\pi$-modes) are depicted in red (black).}
\label{figu4}
\end{figure*}

By following the premise that the position and character of each stationary
point organizes the dynamics, we may conclude from the above study that three topologically
different types of phase portraits can be obtained, depending on which range of values, defined by 
the bifurcation points, the rotation frequency lies. On the other hand, one can identify different regimes
 for each kind of phase portrait.
For instance, we observe in Fig. \ref{figu2} closed orbits around the minimum at the origin,
which are referred to as 0-modes. Analogously, the closed orbits around the minimum at
$\varphi=\pi$ are called $\pi$-modes. Both, the 0- and $\pi$-modes correspond to Josephson
oscillation regimes.
On the other hand, the open orbits with a running phase
not crossing the $Z\equiv 0$ axis, belong to the macroscopic quantum self-trapping regime.
Such a regime is separated from the Josephson regimes by the
limiting orbit known as the separatrix. Such a curve serves as a boundary between different types of orbits
and has been depicted as a thick solid line in Fig. \ref{figu2}.
Since the separatrix has the energy of the saddle points, it is defined by the condition
\begin{equation}
H(Z,\varphi)=H(0,\varphi_s),
\label{hcrit}
\end{equation}
where $H$  denotes the Hamiltonian (\ref{hamilt}).
One can determine from the above equation the maximum
particle imbalance $Z_c$ that can be reached either in the 0- or the $\pi$-modes.
In fact, in the low-frequency interval of Fig. \ref{figu1} ($0<f<f_1$),
 we have only 0-modes ($\varphi=0$) and a saddle at
$\varphi_s=\pi$, which according to (\ref{hcrit}) and (\ref{hamilt}), and
under the assumption that $NU_{\rm eff}$ is much larger than any other model parameter, yields
\begin{equation}
Z_c=\sqrt{\frac{8|K|}{NU_{\rm eff}}},
\label{zc1}
\end{equation}
that is, the same expression obtained for the nonrotating case \cite{nigro17}.
On the other hand, in the central frequency interval between the bifurcation values ($f_1<f<f_2$),
we have both, 0- and $\pi$-modes, i.e., minima at $\varphi=0$ and $\varphi=\pi$, and
saddles at $\varphi_s=\pm\cos^{-1}(K/\varepsilon)$. Thus, replacing such values in Eq. (\ref{hcrit}),
we obtain the following expression for the critical imbalances
$Z_c^+$ and $Z_c^-$ of the 0- and $\pi$-modes, respectively,
\begin{equation}
Z_c^\pm=\sqrt{\frac{-2}{\varepsilon NU_{\rm eff}}(\varepsilon\mp K)^2}.
\label{zcpm}
\end{equation}
Replacing in the above expression the parameter values corresponding to the phase portrait of Fig. \ref{figu2}, one obtains a quotient $Z_c^-/Z_c^+=3$,
which turns out to be in accordance with the separatrix values at $\varphi=\pi$ and 
$\varphi=0$, as seen in the graph.
Finally, it is easy to show that the critical imbalance for the $\pi$-modes of
the last frequency interval, $f>f_2$, in Fig. \ref{figu1},
is again given by the expression (\ref{zc1}). 
We note that it is straightforward to verify that Eqs. (\ref{zc1}) and (\ref{zcpm})
coincide at the bifurcation frequencies, as expected. Such theoretical predictions,
which show an excellent agreement with the GP simulation results, are depicted in Fig. \ref{figu4}.

\section{Dynamics}\label{sec5}
First, it is interesting to analyze the impact that both, the correlated hopping and the intersite interaction parameters,
have on the time periods when varying the rotation frequency.
In the nonrotating case, a system for which such a correction ($\varepsilon=\varepsilon'$)
cannot be neglected was recently considered  \cite{nigro17}. There it was shown that the main effect
was restricted to low imbalances in the Josephson regime. 
That is, such a correction turned out to be appreciable only in the SO limit,
with a decreasing incidence for increasing particle imbalances up to the critical
value and negligible effects in the self-trapping regime. 
In the present rotating case, we will focus on the SO approximation for the time-periods, which 
is expected to be deeply modified at frequencies where
$\varepsilon$ becomes of the order of $K$.
Then, by linearizing the equations of motion we obtain, 
\begin{equation}
T_\pm=\frac{\pi\hbar}{\sqrt{(\pm K-\varepsilon)(NU_{\rm eff}/2\pm K-\varepsilon/2-\varepsilon')}},
\label{period}
\end{equation}
where $T_+$ ($T_-$) denotes the SO period of the 0-($\pi$-)modes. 
We note that the above formula for  $T_+$ reduces to the previously obtained in Ref. \cite{nigro17}
for a nonrotating condensate by taking $\varepsilon=\varepsilon'$.
Given that $NU_{\rm eff}$ turns out to be much greater than the remaining parameters, the intersite interaction has a negligible effect on (\ref{period}). 
On the other hand, divergencies of the period occur at the bifurcation
frequencies due to the $(\pm K-\varepsilon)$ factor in the denominator.
 \begin{figure*}
\includegraphics{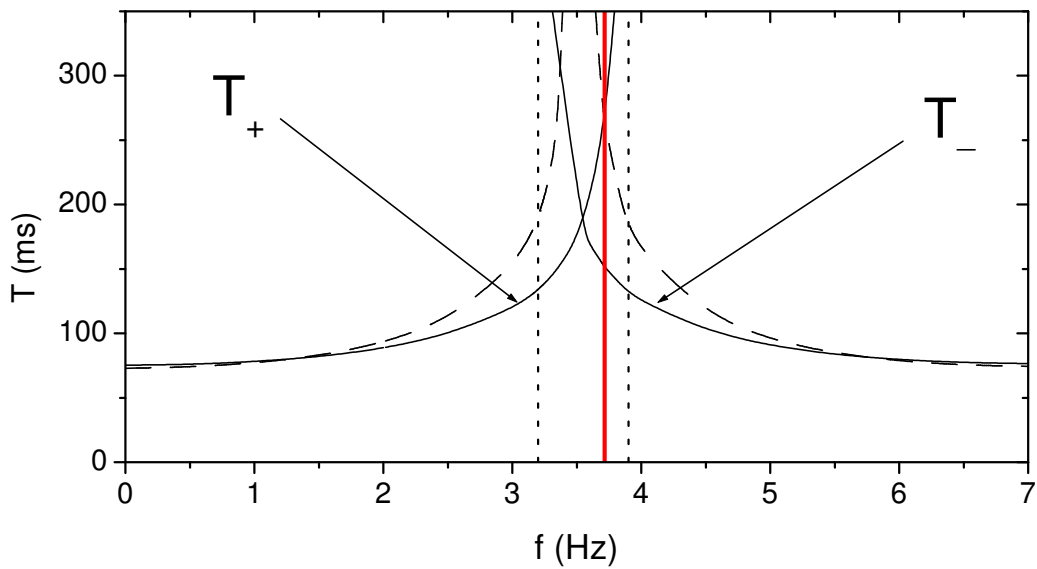}
\caption{SO periods $T_+$ and $T_-$ given by Eq. (\ref{period})
versus the rotational frequency $f$ (solid lines). 
The vertical dotted lines indicate the bifurcation frequencies which
separate the
three frequency intervals of Fig. \ref{figu1}.
Within the central interval, the vertical red line corresponds to the frequency $f=3.72$ Hz
of the phase portrait of Fig. \ref{figu2}.
We also depict as dashed lines the periods obtained from
Eq. (\ref{period}) with vanishing values of $\varepsilon$
and $\varepsilon'$. 
 }
\label{figu3}
\end{figure*}
 In Fig. \ref{figu3}, we display the results arising from Eq. (\ref{period}) (solid lines), together with the corresponding results for vanishing values of $\varepsilon$ and
$\varepsilon'$ (dashed lines). Thus, we may observe
that for the condensate without rotation,
the period does not differ appreciably from that of the $\varepsilon=\varepsilon'=0$
case, as happens for most nonrotating systems.
In contrast, for the rotating condensate,
even out of the two-minimum zone delimited by the bifurcation frequencies,
there exists a sizable difference between the dashed and solid lines in Fig. \ref{figu3}. 
Particularly, above $f=2$ Hz such a
separation between lines becomes very visible, reflecting
the incidence of the correlated hopping on the period. 
 \begin{figure*}
\includegraphics{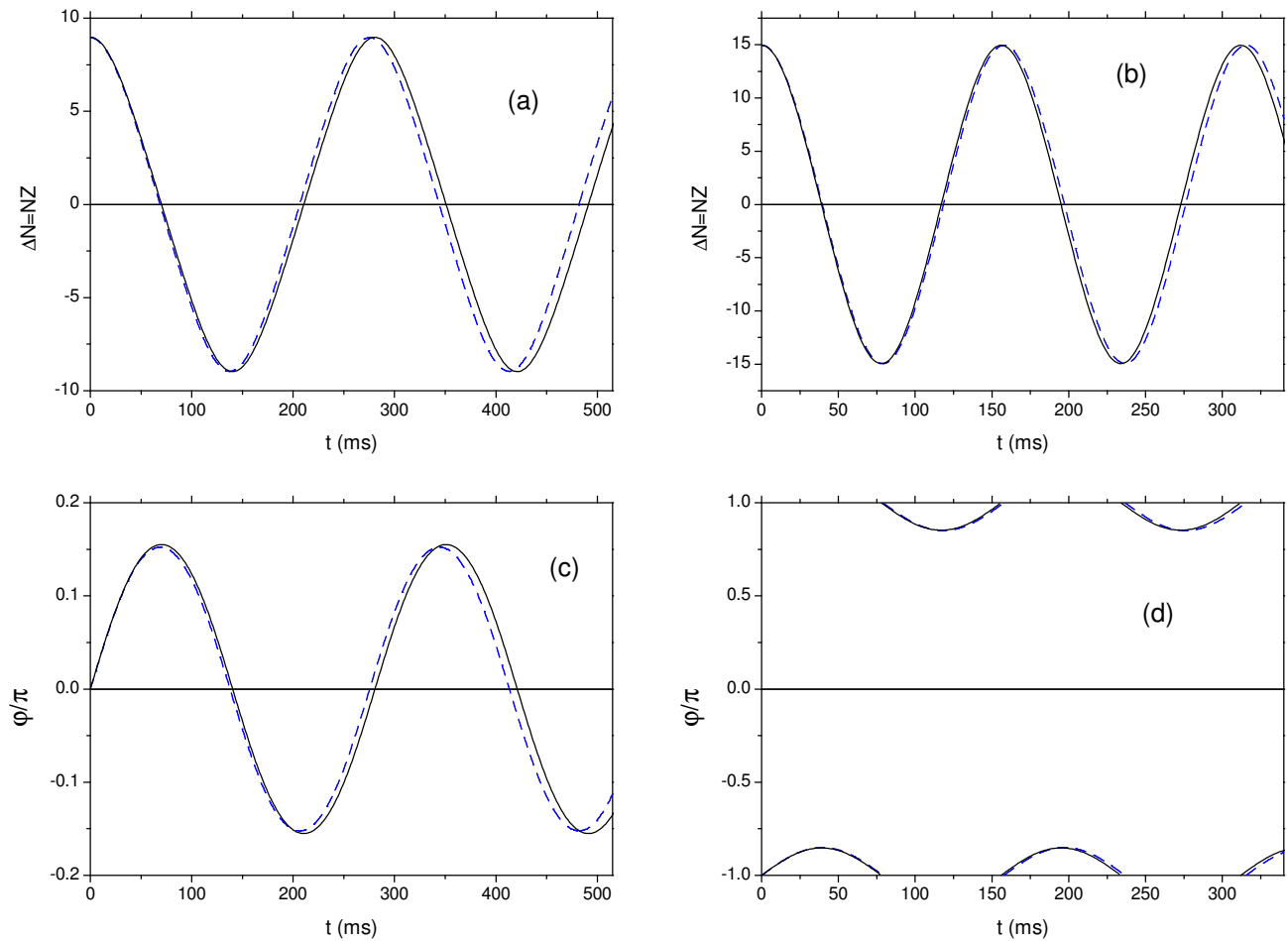}
\caption{Time evolution of the imbalance (panels (a) and (b)) and phase difference 
(panels (c) and (d)) 
for the Josephson modes of the condensate rotating at $f=3.72$ Hz. 
Left ((a) and (c)) and right ((b) and (d)) panels show the 0- and $\pi$-modes, respectively.
Black solid (blue dashed) lines correspond to TM model (GP simulation)
results.}
\label{figu6}
\end{figure*}
 \begin{figure*}
\includegraphics{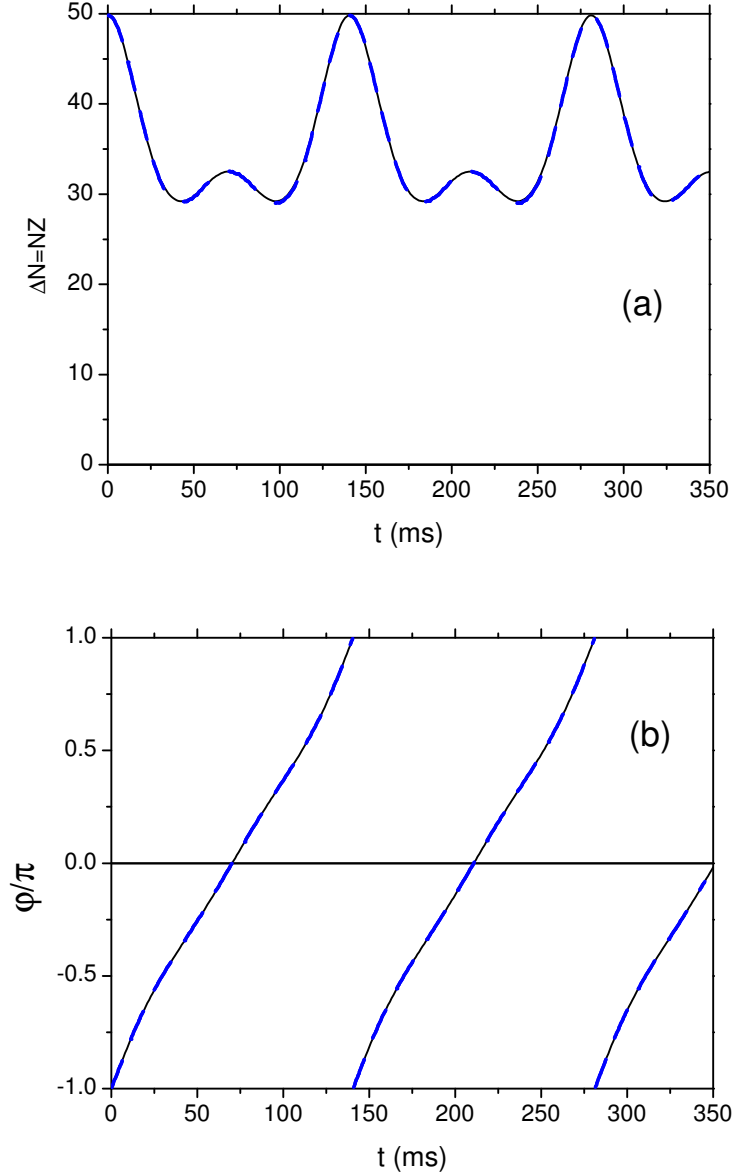}
\caption{Particle imbalance and phase difference versus time are shown in the upper (a) and lower (b) panel, respectively. The initial condition $Z(0)=0.017$ and
 $\varphi(0)=-\pi$ correspond to a self-trapping orbit in Fig. 3. The black solid  and blue dashed lines correspond to  the TM model and GP simulation results, respectively.
}
\label{figu7}
\end{figure*}

Time evolutions of particle imbalance and phase difference for initial conditions within the Josephson and self-trapping  regimes are shown in Figs.~\ref{figu6} and \ref{figu7}, respectively.
We have considered the case of the rotation frequency $3.72$ Hz, the phase space portrait of which was discussed 
in Sec. \ref{sec4}. The left panels of  Fig.~\ref{figu6} show the evolutions obtained from the TM model and the GP simulations for the  $0$-mode Josephson orbit  depicted in Fig. \ref{figu2}, whereas the corresponding evolutions of the $\pi$-mode of the smaller orbit  
are shown on the right panels. 
We note that the time periods in the SO approximation for the present frequency (marked with a vertical red line in Fig. \ref{figu3}) 
are around $ T_{+}\simeq 250$ ms and $T_{-} \simeq 150$ ms, for the $0$- and $\pi$-modes, respectively. Such values can be considered as rough estimates of the periods of Fig.~\ref{figu6}, where we observe
a closer estimate for the $\pi$-mode, despite its larger initial imbalance. This is due to the fact that such an orbit lies farther  from the separatrix than the $0$-mode.   Actually,  it has been shown in Ref. 
\cite{nigro17} that the SO periods constitute  lower bounds of the Josephson periods, which, on the other hand, diverge when approaching the critical imbalance. 

Finally, Fig.~\ref{figu7} displays
the time evolution of particle imbalance and phase
difference for the self-trapping orbit of Fig.~\ref{figu2} with the initial condition $Z(0)=0.017$ and $\varphi(0)=-\pi$.
We may see that the imbalance presents peaks of absolute
and relative maxima corresponding to $\varphi= \pm \pi$ 
and $\varphi= 0 $, respectively, which is in
accordance with having a larger $Z_c$ for the $\pi$-mode than 
that of the $0$-mode, as shown in Fig.~\ref{figu4}.
Given that also in the self-trapping regime the dominant term of the right-hand side of Eq. (\ref{fipun})
is the first one, one can approximate the period as $T_{ST} = 2 \pi \hbar / (\langle Z\rangle N U_{\rm eff}) $, where $\langle Z\rangle$ denotes the mean value of the imbalance over one period. 
Hence, the above modulation  of the imbalance due to the role of the hopping amplitude $\varepsilon$ indirectly affects the time period. In particular, the period observed in Fig.~\ref{figu7} is about 140 ms, which is consistent with $\langle Z\rangle N \simeq 35 $.

To conclude we remark that a very good  agreement between the TM model
and the GP simulation results was observed for all the  rotational frequencies we have explored, and the above exposed results may be taken as a representative example of such an accordance. 
 
\section{Conclusion}\label{sec6}
We have shown that by rotating a toroidal-type double-well condensate, a much richer dynamics of the macroscopic conjugate coordinates can emerge, basically the coexistence of Josephson $0$- and $\pi$-modes.
Our findings have relied on a TM model suitable for such a rotating system. We have shown that the behavior of the hopping amplitudes $K$ and $ \varepsilon $ with respect to rotation leads to an inversion of their relative strengths, as compared to the usual nonrotating relative values. Such parameters exhibit periodic modulations as functions of the rotational frequency, but, given that the period of one of them is  a half of the other, there exist  crossings between  such curves. In particular, at the frequencies where the absolute values of both hopping amplitudes coincide,  bifurcations of stationary points in the phase space occur, which in turn give rise to the coexistence of two regimes of Josephson modes.
We have shown that the frequency periods of such parameters can be easily estimated using a 1D approach, which may be useful for designing  a convenient  experimental set up to reproduce this dynamics.

An extra benefit of having explored the rotation aspects of the system was to gain a more profound insight on the nature of the model parameters.
Due to the phase gradient experimented by the on-site localized functions,
we were able to distinguish between the hopping and interaction parameters $\varepsilon$ and $\varepsilon'$,
which are coincident in the
nonrotating case. Moreover, as previously observed, such a correlated hopping can be related to the tunneling of pairs of particles, which suggests that the coexistence of Josephson modes could be a consequence of such tunneling processes.

To conclude, we have presented a simple and quite accurate bimodal model of the rotation dynamics of a dc-AQUID, which can be regarded as an important advance to be employed to analyze the interference of persistent currents in atomtronic devices.
\appendix*

\section{Alternative calculation of the TM parameters $K$ and $\varepsilon$}
There is a simple relationship between the parameter $K$ and the energy gap between both 
stationary states,
\begin{equation}
K=\Delta E/2,
\label{KD}
\end{equation}
where $\Delta E=E_1-E_0$, being $E_1$ ($E_0$) the energy per particle of the stationary state
 with winding number $n=1$ ($n=0$) obtained from the GP equation.
Equation (\ref{KD}) provides a useful tool to test the accuracy of the TM model.
For instance, if the potential barriers were not high enough, or the on-site functions were not 
properly localized, the energies $E_1$ and $E_0$ would not satisfy (\ref{KD}) \cite{nigro18}.
Then, we have utilized the above formula to test such accuracy  of the $ K$ values obtained from the integrals (\ref{jota}) and (\ref{efe}), finding a coincidence within four digits along the whole frequency range of  Fig. \ref{figu1}. 

It is worthwhile noticing that the above winding numbers correspond to the nonrotating
case, as they can change in amounts of
two units at increasing angular velocities due to the discrete two-fold rotational symmetry of the system
\cite{nigro20}.
For instance, the
stationary state of winding number $n=0$ 
acquires for frequencies approaching 7 Hz the value $n=2$, whereas, the stationary state
of winding number 
$n=1$ conserves this value for the whole frequency range shown in Fig. \ref{figu1}.  

In Sec.~\ref{sec5}, the parameters $\varepsilon$ and $\varepsilon'$
were shown to be related according to Eq.~(\ref{period}) to the SO period arising
from the Hamiltonian (\ref{hamilt}). Now,
taking into account that we work with a $NU_{\rm eff}$ value which is three orders of magnitude larger
than the remaining parameters
$K$, $\varepsilon$ and $\varepsilon'$, we may approximate (\ref{period}) as,
\begin{equation}
T_\pm=\pi\hbar\sqrt{\frac{2}{NU_{\rm eff}(\pm K-\varepsilon)}},
\label{periodapp}
\end{equation}
from where we may obtain the following expression,
\begin{equation}
\varepsilon=\pm K-\frac{2\pi^2\hbar^2}{NU_{\rm eff}T^2_\pm},
\label{varep}
\end{equation}
which was in fact utilized to calculate the values of the parameter $\varepsilon$,
given that it provides results of a much better accuracy than the numerical integration of 
expression (\ref{eps}). 

\acknowledgments
DMJ and HMC  acknowledge CONICET for financial support under Grant
 PIP 11220150100442CO. HMC acknowledge Universidad de Buenos Aires for financial support under
Grant UBA-CyT 20020190100214BA.


\begin{thebibliography}{36}%
\makeatletter
\providecommand \@ifxundefined [1]{%
 \@ifx{#1\undefined}
}%
\providecommand \@ifnum [1]{%
 \ifnum #1\expandafter \@firstoftwo
 \else \expandafter \@secondoftwo
 \fi
}%
\providecommand \@ifx [1]{%
 \ifx #1\expandafter \@firstoftwo
 \else \expandafter \@secondoftwo
 \fi
}%
\providecommand \natexlab [1]{#1}%
\providecommand \enquote  [1]{``#1''}%
\providecommand \bibnamefont  [1]{#1}%
\providecommand \bibfnamefont [1]{#1}%
\providecommand \citenamefont [1]{#1}%
\providecommand \href@noop [0]{\@secondoftwo}%
\providecommand \href [0]{\begingroup \@sanitize@url \@href}%
\providecommand \@href[1]{\@@startlink{#1}\@@href}%
\providecommand \@@href[1]{\endgroup#1\@@endlink}%
\providecommand \@sanitize@url [0]{\catcode `\\12\catcode `\$12\catcode
  `\&12\catcode `\#12\catcode `\^12\catcode `\_12\catcode `\%12\relax}%
\providecommand \@@startlink[1]{}%
\providecommand \@@endlink[0]{}%
\providecommand \url  [0]{\begingroup\@sanitize@url \@url }%
\providecommand \@url [1]{\endgroup\@href {#1}{\urlprefix }}%
\providecommand \urlprefix  [0]{URL }%
\providecommand \Eprint [0]{\href }%
\providecommand \doibase [0]{https://doi.org/}%
\providecommand \selectlanguage [0]{\@gobble}%
\providecommand \bibinfo  [0]{\@secondoftwo}%
\providecommand \bibfield  [0]{\@secondoftwo}%
\providecommand \translation [1]{[#1]}%
\providecommand \BibitemOpen [0]{}%
\providecommand \bibitemStop [0]{}%
\providecommand \bibitemNoStop [0]{.\EOS\space}%
\providecommand \EOS [0]{\spacefactor3000\relax}%
\providecommand \BibitemShut  [1]{\csname bibitem#1\endcsname}%
\let\auto@bib@innerbib\@empty
\bibitem [{\citenamefont {Amico}\ \emph {et~al.}(2020)\citenamefont {Amico},
  \citenamefont {Boshier}, \citenamefont {Birkl}, \citenamefont {Minguzzi},
  \citenamefont {Miniatura}, \citenamefont {{\relax L. -C. Kwek}},
  \citenamefont {Aghamalyan}, \citenamefont {Ahufinger}, \citenamefont
  {Anderson}, \citenamefont {Andrei} \emph {et~al.}}]{amico2021state}%
  \BibitemOpen
  \bibfield  {author} {\bibinfo {author} {\bibfnamefont {L.}~\bibnamefont
  {Amico}}, \bibinfo {author} {\bibfnamefont {M.}~\bibnamefont {Boshier}},
  \bibinfo {author} {\bibfnamefont {G.}~\bibnamefont {Birkl}}, \bibinfo
  {author} {\bibfnamefont {A.}~\bibnamefont {Minguzzi}}, \bibinfo {author}
  {\bibfnamefont {C.}~\bibnamefont {Miniatura}}, \bibinfo {author}
  {\bibnamefont {{\relax L. -C. Kwek}}}, \bibinfo {author} {\bibfnamefont
  {D.}~\bibnamefont {Aghamalyan}}, \bibinfo {author} {\bibfnamefont
  {V.}~\bibnamefont {Ahufinger}}, \bibinfo {author} {\bibfnamefont
  {D.}~\bibnamefont {Anderson}}, \bibinfo {author} {\bibfnamefont
  {N.}~\bibnamefont {Andrei}}, \emph {et~al.},\ }\href@noop {} {} (\bibinfo
  {year} {2020}),\ \Eprint {https://arxiv.org/abs/2008.04439} {arXiv:2008.04439
  [cond-mat.quant-gas]} \BibitemShut {NoStop}%
\bibitem [{\citenamefont {Amico}\ \emph {et~al.}(2021)\citenamefont {Amico},
  \citenamefont {Anderson}, \citenamefont {Boshier}, \citenamefont {{\relax J.
  -P. Brantut}}, \citenamefont {{\relax L. -C. Kwek}}, \citenamefont
  {Minguzzi},\ and\ \citenamefont {{\relax W. von Klitzing}}}]{acircuits}%
  \BibitemOpen
  \bibfield  {author} {\bibinfo {author} {\bibfnamefont {L.}~\bibnamefont
  {Amico}}, \bibinfo {author} {\bibfnamefont {D.}~\bibnamefont {Anderson}},
  \bibinfo {author} {\bibfnamefont {M.}~\bibnamefont {Boshier}}, \bibinfo
  {author} {\bibnamefont {{\relax J. -P. Brantut}}}, \bibinfo {author}
  {\bibnamefont {{\relax L. -C. Kwek}}}, \bibinfo {author} {\bibfnamefont
  {A.}~\bibnamefont {Minguzzi}},\ and\ \bibinfo {author} {\bibnamefont {{\relax
  W. von Klitzing}}},\ }\href@noop {} {} (\bibinfo {year} {2021}),\ \Eprint
  {https://arxiv.org/abs/2107.08561} {arXiv:2107.08561 [cond-mat.quant-gas]}
  \BibitemShut {NoStop}%
\bibitem [{\citenamefont {Clarke}\ and\ \citenamefont
  {Braginski}(2004)}]{braginski}%
  \BibitemOpen
  \bibfield  {author} {\bibinfo {author} {\bibfnamefont {J.}~\bibnamefont
  {Clarke}}\ and\ \bibinfo {author} {\bibfnamefont {A.~I.}\ \bibnamefont
  {Braginski}},\ }\href@noop {} {\emph {\bibinfo {title} {The SQUID
  Handbook}}}\ (\bibinfo  {publisher} {Wiley-VCH},\ \bibinfo {address}
  {Weinheim},\ \bibinfo {year} {2004})\BibitemShut {NoStop}%
\bibitem [{\citenamefont {Fagaly}(2006)}]{fagaly}%
  \BibitemOpen
  \bibfield  {author} {\bibinfo {author} {\bibfnamefont {R.~L.}\ \bibnamefont
  {Fagaly}},\ }\href@noop {} {\bibfield  {journal} {\bibinfo  {journal} {Rev.
  Sci. Instrum.}\ }\textbf {\bibinfo {volume} {77}},\ \bibinfo {pages} {101101}
  (\bibinfo {year} {2006})}\BibitemShut {NoStop}%
\bibitem [{\citenamefont {Wright}\ \emph {et~al.}(2013)\citenamefont {Wright},
  \citenamefont {Blakestad}, \citenamefont {Lobb}, \citenamefont {Phillips},\
  and\ \citenamefont {Campbell}}]{wright}%
  \BibitemOpen
  \bibfield  {author} {\bibinfo {author} {\bibfnamefont {K.~C.}\ \bibnamefont
  {Wright}}, \bibinfo {author} {\bibfnamefont {R.~B.}\ \bibnamefont
  {Blakestad}}, \bibinfo {author} {\bibfnamefont {C.~J.}\ \bibnamefont {Lobb}},
  \bibinfo {author} {\bibfnamefont {W.~D.}\ \bibnamefont {Phillips}},\ and\
  \bibinfo {author} {\bibfnamefont {G.~K.}\ \bibnamefont {Campbell}},\
  }\href@noop {} {\bibfield  {journal} {\bibinfo  {journal} {Phys. Rev. Lett.}\
  }\textbf {\bibinfo {volume} {110}},\ \bibinfo {pages} {025302} (\bibinfo
  {year} {2013})}\BibitemShut {NoStop}%
\bibitem [{\citenamefont {Eckel}\ \emph {et~al.}(2014)\citenamefont {Eckel},
  \citenamefont {Lee}, \citenamefont {Jendrzejewski}, \citenamefont {Murray},
  \citenamefont {Clark}, \citenamefont {Lobb}, \citenamefont {Phillips},
  \citenamefont {Edwards},\ and\ \citenamefont {Campbell}}]{eckel}%
  \BibitemOpen
  \bibfield  {author} {\bibinfo {author} {\bibfnamefont {S.}~\bibnamefont
  {Eckel}}, \bibinfo {author} {\bibfnamefont {J.~G.}\ \bibnamefont {Lee}},
  \bibinfo {author} {\bibfnamefont {F.}~\bibnamefont {Jendrzejewski}}, \bibinfo
  {author} {\bibfnamefont {N.}~\bibnamefont {Murray}}, \bibinfo {author}
  {\bibfnamefont {C.~W.}\ \bibnamefont {Clark}}, \bibinfo {author}
  {\bibfnamefont {C.~J.}\ \bibnamefont {Lobb}}, \bibinfo {author}
  {\bibfnamefont {W.~D.}\ \bibnamefont {Phillips}}, \bibinfo {author}
  {\bibfnamefont {M.}~\bibnamefont {Edwards}},\ and\ \bibinfo {author}
  {\bibfnamefont {G.~K.}\ \bibnamefont {Campbell}},\ }\href
  {https://doi.org/10.1038/nature12958} {\bibfield  {journal} {\bibinfo
  {journal} {Nature}\ }\textbf {\bibinfo {volume} {506}},\ \bibinfo {pages}
  {200} (\bibinfo {year} {2014})}\BibitemShut {NoStop}%
\bibitem [{\citenamefont {Dalibard}\ \emph {et~al.}(2011)\citenamefont
  {Dalibard}, \citenamefont {Gerbier}, \citenamefont {Juzeliunas},\ and\
  \citenamefont {\"Ohberg}}]{rmp}%
  \BibitemOpen
  \bibfield  {author} {\bibinfo {author} {\bibfnamefont {J.}~\bibnamefont
  {Dalibard}}, \bibinfo {author} {\bibfnamefont {F.}~\bibnamefont {Gerbier}},
  \bibinfo {author} {\bibfnamefont {G.}~\bibnamefont {Juzeliunas}},\ and\
  \bibinfo {author} {\bibfnamefont {P.}~\bibnamefont {\"Ohberg}},\ }\href@noop
  {} {\bibfield  {journal} {\bibinfo  {journal} {Rev. Mod. Phys.}\ }\textbf
  {\bibinfo {volume} {83}},\ \bibinfo {pages} {1523} (\bibinfo {year}
  {2011})}\BibitemShut {NoStop}%
\bibitem [{\citenamefont {Sato}\ and\ \citenamefont {Packard}(2012)}]{satop}%
  \BibitemOpen
  \bibfield  {author} {\bibinfo {author} {\bibfnamefont {Y.}~\bibnamefont
  {Sato}}\ and\ \bibinfo {author} {\bibfnamefont {R.~E.}\ \bibnamefont
  {Packard}},\ }\href@noop {} {\bibfield  {journal} {\bibinfo  {journal} {Rep.
  Prog. Phys.}\ }\textbf {\bibinfo {volume} {75}},\ \bibinfo {pages} {016401}
  (\bibinfo {year} {2012})}\BibitemShut {NoStop}%
\bibitem [{\citenamefont {Ryu}\ \emph {et~al.}(2020)\citenamefont {Ryu},
  \citenamefont {Samson},\ and\ \citenamefont {Boshier}}]{ryu20}%
  \BibitemOpen
  \bibfield  {author} {\bibinfo {author} {\bibfnamefont {C.}~\bibnamefont
  {Ryu}}, \bibinfo {author} {\bibfnamefont {E.~C.}\ \bibnamefont {Samson}},\
  and\ \bibinfo {author} {\bibfnamefont {M.~G.}\ \bibnamefont {Boshier}},\
  }\href@noop {} {\bibfield  {journal} {\bibinfo  {journal} {Nat. Commun.}\
  }\textbf {\bibinfo {volume} {11}},\ \bibinfo {pages} {3338} (\bibinfo {year}
  {2020})}\BibitemShut {NoStop}%
\bibitem [{\citenamefont {{\relax G. Pelegr\'{\i}}}\ \emph
  {et~al.}(2018)\citenamefont {{\relax G. Pelegr\'{\i}}}, \citenamefont
  {Mompart},\ and\ \citenamefont {Ahufinger}}]{pele}%
  \BibitemOpen
  \bibfield  {author} {\bibinfo {author} {\bibnamefont {{\relax G.
  Pelegr\'{\i}}}}, \bibinfo {author} {\bibfnamefont {J.}~\bibnamefont
  {Mompart}},\ and\ \bibinfo {author} {\bibfnamefont {V.}~\bibnamefont
  {Ahufinger}},\ }\href@noop {} {\bibfield  {journal} {\bibinfo  {journal} {New
  J. Phys.}\ }\textbf {\bibinfo {volume} {20}},\ \bibinfo {pages} {103001}
  (\bibinfo {year} {2018})}\BibitemShut {NoStop}%
\bibitem [{\citenamefont {Nicolau}\ \emph {et~al.}(2020)\citenamefont
  {Nicolau}, \citenamefont {Mompart}, \citenamefont {{\relax B.
  Juli\'a-D\'{\i}az}},\ and\ \citenamefont {Ahufinger}}]{nico}%
  \BibitemOpen
  \bibfield  {author} {\bibinfo {author} {\bibfnamefont {E.}~\bibnamefont
  {Nicolau}}, \bibinfo {author} {\bibfnamefont {J.}~\bibnamefont {Mompart}},
  \bibinfo {author} {\bibnamefont {{\relax B. Juli\'a-D\'{\i}az}}},\ and\
  \bibinfo {author} {\bibfnamefont {V.}~\bibnamefont {Ahufinger}},\ }\href@noop
  {} {\bibfield  {journal} {\bibinfo  {journal} {Phys. Rev. A}\ }\textbf
  {\bibinfo {volume} {102}},\ \bibinfo {pages} {023331} (\bibinfo {year}
  {2020})}\BibitemShut {NoStop}%
\bibitem [{\citenamefont {Kumar}\ \emph {et~al.}(2021)\citenamefont {Kumar},
  \citenamefont {Biswas}, \citenamefont {Feliz}, \citenamefont {Kanamoto},
  \citenamefont {{\relax M.-S. Chang}}, \citenamefont {Jha},\ and\
  \citenamefont {Bhattacharya}}]{kumar}%
  \BibitemOpen
  \bibfield  {author} {\bibinfo {author} {\bibfnamefont {P.}~\bibnamefont
  {Kumar}}, \bibinfo {author} {\bibfnamefont {T.}~\bibnamefont {Biswas}},
  \bibinfo {author} {\bibfnamefont {K.}~\bibnamefont {Feliz}}, \bibinfo
  {author} {\bibfnamefont {R.}~\bibnamefont {Kanamoto}}, \bibinfo {author}
  {\bibnamefont {{\relax M.-S. Chang}}}, \bibinfo {author} {\bibfnamefont
  {A.~K.}\ \bibnamefont {Jha}},\ and\ \bibinfo {author} {\bibfnamefont
  {M.}~\bibnamefont {Bhattacharya}},\ }\href@noop {} {\bibfield  {journal}
  {\bibinfo  {journal} {Phys. Rev. Lett.}\ }\textbf {\bibinfo {volume} {127}},\
  \bibinfo {pages} {113601} (\bibinfo {year} {2021})}\BibitemShut {NoStop}%
\bibitem [{\citenamefont {Degen}\ \emph {et~al.}(2017)\citenamefont {Degen},
  \citenamefont {Reinhard},\ and\ \citenamefont {Cappellaro}}]{degen}%
  \BibitemOpen
  \bibfield  {author} {\bibinfo {author} {\bibfnamefont {C.~L.}\ \bibnamefont
  {Degen}}, \bibinfo {author} {\bibfnamefont {F.}~\bibnamefont {Reinhard}},\
  and\ \bibinfo {author} {\bibfnamefont {P.}~\bibnamefont {Cappellaro}},\
  }\href@noop {} {\bibfield  {journal} {\bibinfo  {journal} {Rev. Mod. Phys.}\
  }\textbf {\bibinfo {volume} {89}},\ \bibinfo {pages} {035002} (\bibinfo
  {year} {2017})}\BibitemShut {NoStop}%
\bibitem [{\citenamefont {Ryu}\ \emph {et~al.}(2013)\citenamefont {Ryu},
  \citenamefont {Blackburn}, \citenamefont {Blinova},\ and\ \citenamefont
  {Boshier}}]{boshier13}%
  \BibitemOpen
  \bibfield  {author} {\bibinfo {author} {\bibfnamefont {C.}~\bibnamefont
  {Ryu}}, \bibinfo {author} {\bibfnamefont {P.~W.}\ \bibnamefont {Blackburn}},
  \bibinfo {author} {\bibfnamefont {A.~A.}\ \bibnamefont {Blinova}},\ and\
  \bibinfo {author} {\bibfnamefont {M.~G.}\ \bibnamefont {Boshier}},\
  }\href@noop {} {\bibfield  {journal} {\bibinfo  {journal} {Phys. Rev. Lett.}\
  }\textbf {\bibinfo {volume} {111}},\ \bibinfo {pages} {205301} (\bibinfo
  {year} {2013})}\BibitemShut {NoStop}%
\bibitem [{\citenamefont {Jendrzejewski}\ \emph {et~al.}(2014)\citenamefont
  {Jendrzejewski}, \citenamefont {Eckel}, \citenamefont {Murray}, \citenamefont
  {Lanier}, \citenamefont {Edwards}, \citenamefont {Lobb},\ and\ \citenamefont
  {Campbell}}]{jen14}%
  \BibitemOpen
  \bibfield  {author} {\bibinfo {author} {\bibfnamefont {F.}~\bibnamefont
  {Jendrzejewski}}, \bibinfo {author} {\bibfnamefont {S.}~\bibnamefont
  {Eckel}}, \bibinfo {author} {\bibfnamefont {N.}~\bibnamefont {Murray}},
  \bibinfo {author} {\bibfnamefont {C.}~\bibnamefont {Lanier}}, \bibinfo
  {author} {\bibfnamefont {M.}~\bibnamefont {Edwards}}, \bibinfo {author}
  {\bibfnamefont {C.~J.}\ \bibnamefont {Lobb}},\ and\ \bibinfo {author}
  {\bibfnamefont {G.~K.}\ \bibnamefont {Campbell}},\ }\href@noop {} {\bibfield
  {journal} {\bibinfo  {journal} {Phys. Rev. Lett.}\ }\textbf {\bibinfo
  {volume} {113}},\ \bibinfo {pages} {045305} (\bibinfo {year}
  {2014})}\BibitemShut {NoStop}%
\bibitem [{\citenamefont {Aghalmalyan}\ \emph {et~al.}(2015)\citenamefont
  {Aghalmalyan}, \citenamefont {Cominotti}, \citenamefont {Rizzi},
  \citenamefont {Rossini}, \citenamefont {Hekking}, \citenamefont {Minguzzi},
  \citenamefont {{\relax L. -C. Kwek}},\ and\ \citenamefont
  {Amico}}]{rf-aquid}%
  \BibitemOpen
  \bibfield  {author} {\bibinfo {author} {\bibfnamefont {D.}~\bibnamefont
  {Aghalmalyan}}, \bibinfo {author} {\bibfnamefont {M.}~\bibnamefont
  {Cominotti}}, \bibinfo {author} {\bibfnamefont {M.}~\bibnamefont {Rizzi}},
  \bibinfo {author} {\bibfnamefont {D.}~\bibnamefont {Rossini}}, \bibinfo
  {author} {\bibfnamefont {F.}~\bibnamefont {Hekking}}, \bibinfo {author}
  {\bibfnamefont {A.}~\bibnamefont {Minguzzi}}, \bibinfo {author} {\bibnamefont
  {{\relax L. -C. Kwek}}},\ and\ \bibinfo {author} {\bibfnamefont
  {L.}~\bibnamefont {Amico}},\ }\href@noop {} {\bibfield  {journal} {\bibinfo
  {journal} {New J. Phys.}\ }\textbf {\bibinfo {volume} {17}},\ \bibinfo
  {pages} {045023} (\bibinfo {year} {2015})}\BibitemShut {NoStop}%
\bibitem [{\citenamefont {Aghalmalyan}\ \emph {et~al.}(2016)\citenamefont
  {Aghalmalyan}, \citenamefont {Nguyen}, \citenamefont {Auksztol},
  \citenamefont {Gan}, \citenamefont {{\relax M. Mart\'{\i}nez Valado}},
  \citenamefont {Condylis}, \citenamefont {{\relax L. -C. Kwek}}, \citenamefont
  {Dumke},\ and\ \citenamefont {Amico}}]{3-wl}%
  \BibitemOpen
  \bibfield  {author} {\bibinfo {author} {\bibfnamefont {D.}~\bibnamefont
  {Aghalmalyan}}, \bibinfo {author} {\bibfnamefont {N.~T.}\ \bibnamefont
  {Nguyen}}, \bibinfo {author} {\bibfnamefont {F.}~\bibnamefont {Auksztol}},
  \bibinfo {author} {\bibfnamefont {K.~S.}\ \bibnamefont {Gan}}, \bibinfo
  {author} {\bibnamefont {{\relax M. Mart\'{\i}nez Valado}}}, \bibinfo {author}
  {\bibfnamefont {P.~C.}\ \bibnamefont {Condylis}}, \bibinfo {author}
  {\bibnamefont {{\relax L. -C. Kwek}}}, \bibinfo {author} {\bibfnamefont
  {R.}~\bibnamefont {Dumke}},\ and\ \bibinfo {author} {\bibfnamefont
  {L.}~\bibnamefont {Amico}},\ }\href@noop {} {\bibfield  {journal} {\bibinfo
  {journal} {New J. Phys.}\ }\textbf {\bibinfo {volume} {18}},\ \bibinfo
  {pages} {075013} (\bibinfo {year} {2016})}\BibitemShut {NoStop}%
\bibitem [{\citenamefont {Amico}\ \emph {et~al.}(2014)\citenamefont {Amico},
  \citenamefont {Aghalmalyan}, \citenamefont {Auksztol}, \citenamefont
  {Crepaz}, \citenamefont {Dumke},\ and\ \citenamefont {{\relax L. -C.
  Kwek}}}]{amico14}%
  \BibitemOpen
  \bibfield  {author} {\bibinfo {author} {\bibfnamefont {L.}~\bibnamefont
  {Amico}}, \bibinfo {author} {\bibfnamefont {D.}~\bibnamefont {Aghalmalyan}},
  \bibinfo {author} {\bibfnamefont {F.}~\bibnamefont {Auksztol}}, \bibinfo
  {author} {\bibfnamefont {H.}~\bibnamefont {Crepaz}}, \bibinfo {author}
  {\bibfnamefont {R.}~\bibnamefont {Dumke}},\ and\ \bibinfo {author}
  {\bibnamefont {{\relax L. -C. Kwek}}},\ }\href@noop {} {\bibfield  {journal}
  {\bibinfo  {journal} {Sci. Rep.}\ }\textbf {\bibinfo {volume} {4}},\ \bibinfo
  {pages} {4298} (\bibinfo {year} {2014})}\BibitemShut {NoStop}%
\bibitem [{\citenamefont {Cazalilla}\ \emph {et~al.}(2011)\citenamefont
  {Cazalilla}, \citenamefont {Citro}, \citenamefont {Giamarchi}, \citenamefont
  {Orignac},\ and\ \citenamefont {Rigol}}]{caza}%
  \BibitemOpen
  \bibfield  {author} {\bibinfo {author} {\bibfnamefont {M.~A.}\ \bibnamefont
  {Cazalilla}}, \bibinfo {author} {\bibfnamefont {R.}~\bibnamefont {Citro}},
  \bibinfo {author} {\bibfnamefont {T.}~\bibnamefont {Giamarchi}}, \bibinfo
  {author} {\bibfnamefont {E.}~\bibnamefont {Orignac}},\ and\ \bibinfo {author}
  {\bibfnamefont {M.}~\bibnamefont {Rigol}},\ }\href@noop {} {\bibfield
  {journal} {\bibinfo  {journal} {Rev. Mod. Phys.}\ }\textbf {\bibinfo {volume}
  {83}},\ \bibinfo {pages} {1405} (\bibinfo {year} {2011})}\BibitemShut
  {NoStop}%
\bibitem [{\citenamefont {Polo}\ \emph {et~al.}(2019)\citenamefont {Polo},
  \citenamefont {Dubessy}, \citenamefont {Pedri}, \citenamefont {Perrin},\ and\
  \citenamefont {Minguzzi}}]{polo}%
  \BibitemOpen
  \bibfield  {author} {\bibinfo {author} {\bibfnamefont {J.}~\bibnamefont
  {Polo}}, \bibinfo {author} {\bibfnamefont {R.}~\bibnamefont {Dubessy}},
  \bibinfo {author} {\bibfnamefont {P.}~\bibnamefont {Pedri}}, \bibinfo
  {author} {\bibfnamefont {H.}~\bibnamefont {Perrin}},\ and\ \bibinfo {author}
  {\bibfnamefont {A.}~\bibnamefont {Minguzzi}},\ }\href@noop {} {\bibfield
  {journal} {\bibinfo  {journal} {Phys. Rev. Lett.}\ }\textbf {\bibinfo
  {volume} {123}},\ \bibinfo {pages} {195301} (\bibinfo {year}
  {2019})}\BibitemShut {NoStop}%
\bibitem [{\citenamefont {{\relax A. P\'erez-Obiol}}\ \emph
  {et~al.}(2020)\citenamefont {{\relax A. P\'erez-Obiol}}, \citenamefont
  {Polo},\ and\ \citenamefont {Cheon}}]{polo1}%
  \BibitemOpen
  \bibfield  {author} {\bibinfo {author} {\bibnamefont {{\relax A.
  P\'erez-Obiol}}}, \bibinfo {author} {\bibfnamefont {J.}~\bibnamefont
  {Polo}},\ and\ \bibinfo {author} {\bibfnamefont {T.}~\bibnamefont {Cheon}},\
  }\href@noop {} {\bibfield  {journal} {\bibinfo  {journal} {Phys. Rev. A}\
  }\textbf {\bibinfo {volume} {102}},\ \bibinfo {pages} {063302} (\bibinfo
  {year} {2020})}\BibitemShut {NoStop}%
\bibitem [{\citenamefont {Fazio}\ and\ \citenamefont {{\relax H. van der
  Zant}}(2001)}]{fazio}%
  \BibitemOpen
  \bibfield  {author} {\bibinfo {author} {\bibfnamefont {R.}~\bibnamefont
  {Fazio}}\ and\ \bibinfo {author} {\bibnamefont {{\relax H. van der Zant}}},\
  }\href@noop {} {\bibfield  {journal} {\bibinfo  {journal} {Phys. Rep.}\
  }\textbf {\bibinfo {volume} {355}},\ \bibinfo {pages} {235} (\bibinfo {year}
  {2001})}\BibitemShut {NoStop}%
\bibitem [{\citenamefont {Smerzi}\ \emph {et~al.}(1997)\citenamefont {Smerzi},
  \citenamefont {Fantoni}, \citenamefont {Giovanazzi},\ and\ \citenamefont
  {Shenoy}}]{smerzi97}%
  \BibitemOpen
  \bibfield  {author} {\bibinfo {author} {\bibfnamefont {A.}~\bibnamefont
  {Smerzi}}, \bibinfo {author} {\bibfnamefont {S.}~\bibnamefont {Fantoni}},
  \bibinfo {author} {\bibfnamefont {S.}~\bibnamefont {Giovanazzi}},\ and\
  \bibinfo {author} {\bibfnamefont {S.~R.}\ \bibnamefont {Shenoy}},\
  }\href@noop {} {\bibfield  {journal} {\bibinfo  {journal} {Phys. Rev. Lett.}\
  }\textbf {\bibinfo {volume} {79}},\ \bibinfo {pages} {4950} (\bibinfo {year}
  {1997})}\BibitemShut {NoStop}%
\bibitem [{\citenamefont {Raghavan}\ \emph {et~al.}(1999)\citenamefont
  {Raghavan}, \citenamefont {Smerzi}, \citenamefont {Fantoni},\ and\
  \citenamefont {Shenoy}}]{ragh99}%
  \BibitemOpen
  \bibfield  {author} {\bibinfo {author} {\bibfnamefont {S.}~\bibnamefont
  {Raghavan}}, \bibinfo {author} {\bibfnamefont {A.}~\bibnamefont {Smerzi}},
  \bibinfo {author} {\bibfnamefont {S.}~\bibnamefont {Fantoni}},\ and\ \bibinfo
  {author} {\bibfnamefont {S.~R.}\ \bibnamefont {Shenoy}},\ }\href@noop {}
  {\bibfield  {journal} {\bibinfo  {journal} {Phys. Rev. A}\ }\textbf {\bibinfo
  {volume} {59}},\ \bibinfo {pages} {620} (\bibinfo {year} {1999})}\BibitemShut
  {NoStop}%
\bibitem [{\citenamefont {Ananikian}\ and\ \citenamefont
  {Bergeman}(2006)}]{anan06}%
  \BibitemOpen
  \bibfield  {author} {\bibinfo {author} {\bibfnamefont {D.}~\bibnamefont
  {Ananikian}}\ and\ \bibinfo {author} {\bibfnamefont {T.}~\bibnamefont
  {Bergeman}},\ }\href@noop {} {\bibfield  {journal} {\bibinfo  {journal}
  {Phys. Rev. A}\ }\textbf {\bibinfo {volume} {73}},\ \bibinfo {pages} {013604}
  (\bibinfo {year} {2006})}\BibitemShut {NoStop}%
\bibitem [{\citenamefont {Jezek}\ \emph {et~al.}(2013)\citenamefont {Jezek},
  \citenamefont {Capuzzi},\ and\ \citenamefont {Cataldo}}]{cap13}%
  \BibitemOpen
  \bibfield  {author} {\bibinfo {author} {\bibfnamefont {D.~M.}\ \bibnamefont
  {Jezek}}, \bibinfo {author} {\bibfnamefont {P.}~\bibnamefont {Capuzzi}},\
  and\ \bibinfo {author} {\bibfnamefont {H.~M.}\ \bibnamefont {Cataldo}},\
  }\href@noop {} {\bibfield  {journal} {\bibinfo  {journal} {Phys. Rev. A}\
  }\textbf {\bibinfo {volume} {87}},\ \bibinfo {pages} {053625} (\bibinfo
  {year} {2013})}\BibitemShut {NoStop}%
\bibitem [{\citenamefont {Nigro}\ \emph {et~al.}(2017)\citenamefont {Nigro},
  \citenamefont {Capuzzi}, \citenamefont {Cataldo},\ and\ \citenamefont
  {Jezek}}]{nigro17}%
  \BibitemOpen
  \bibfield  {author} {\bibinfo {author} {\bibfnamefont {M.}~\bibnamefont
  {Nigro}}, \bibinfo {author} {\bibfnamefont {P.}~\bibnamefont {Capuzzi}},
  \bibinfo {author} {\bibfnamefont {H.~M.}\ \bibnamefont {Cataldo}},\ and\
  \bibinfo {author} {\bibfnamefont {D.~M.}\ \bibnamefont {Jezek}},\ }\href@noop
  {} {\bibfield  {journal} {\bibinfo  {journal} {Eur. Phys. J. D}\ }\textbf
  {\bibinfo {volume} {71}},\ \bibinfo {pages} {297} (\bibinfo {year}
  {2017})}\BibitemShut {NoStop}%
\bibitem [{\citenamefont {Cataldo}\ and\ \citenamefont {Jezek}(2014)}]{cat14}%
  \BibitemOpen
  \bibfield  {author} {\bibinfo {author} {\bibfnamefont {H.~M.}\ \bibnamefont
  {Cataldo}}\ and\ \bibinfo {author} {\bibfnamefont {D.~M.}\ \bibnamefont
  {Jezek}},\ }\href@noop {} {\bibfield  {journal} {\bibinfo  {journal} {Phys.
  Rev. A}\ }\textbf {\bibinfo {volume} {90}},\ \bibinfo {pages} {043610}
  (\bibinfo {year} {2014})}\BibitemShut {NoStop}%
\bibitem [{\citenamefont {Cataldo}(2020)}]{cat20}%
  \BibitemOpen
  \bibfield  {author} {\bibinfo {author} {\bibfnamefont {H.~M.}\ \bibnamefont
  {Cataldo}},\ }\href@noop {} {\bibfield  {journal} {\bibinfo  {journal} {Phys.
  Rev. A}\ }\textbf {\bibinfo {volume} {102}},\ \bibinfo {pages} {023323}
  (\bibinfo {year} {2020})}\BibitemShut {NoStop}%
\bibitem [{\citenamefont {Jezek}\ and\ \citenamefont {Cataldo}(2013)}]{je13}%
  \BibitemOpen
  \bibfield  {author} {\bibinfo {author} {\bibfnamefont {D.~M.}\ \bibnamefont
  {Jezek}}\ and\ \bibinfo {author} {\bibfnamefont {H.~M.}\ \bibnamefont
  {Cataldo}},\ }\href@noop {} {\bibfield  {journal} {\bibinfo  {journal} {Phys.
  Rev. A}\ }\textbf {\bibinfo {volume} {88}},\ \bibinfo {pages} {013636}
  (\bibinfo {year} {2013})}\BibitemShut {NoStop}%
\bibitem [{\citenamefont {Nigro}\ \emph {et~al.}(2018)\citenamefont {Nigro},
  \citenamefont {Capuzzi}, \citenamefont {Cataldo},\ and\ \citenamefont
  {Jezek}}]{nigro18}%
  \BibitemOpen
  \bibfield  {author} {\bibinfo {author} {\bibfnamefont {M.}~\bibnamefont
  {Nigro}}, \bibinfo {author} {\bibfnamefont {P.}~\bibnamefont {Capuzzi}},
  \bibinfo {author} {\bibfnamefont {H.~M.}\ \bibnamefont {Cataldo}},\ and\
  \bibinfo {author} {\bibfnamefont {D.~M.}\ \bibnamefont {Jezek}},\ }\href@noop
  {} {\bibfield  {journal} {\bibinfo  {journal} {Phys. Rev. A}\ }\textbf
  {\bibinfo {volume} {97}},\ \bibinfo {pages} {013626} (\bibinfo {year}
  {2018})}\BibitemShut {NoStop}%
\bibitem [{\citenamefont {Nigro}\ \emph {et~al.}(2020)\citenamefont {Nigro},
  \citenamefont {Capuzzi},\ and\ \citenamefont {Jezek}}]{nigro20}%
  \BibitemOpen
  \bibfield  {author} {\bibinfo {author} {\bibfnamefont {M.}~\bibnamefont
  {Nigro}}, \bibinfo {author} {\bibfnamefont {P.}~\bibnamefont {Capuzzi}},\
  and\ \bibinfo {author} {\bibfnamefont {D.~M.}\ \bibnamefont {Jezek}},\
  }\href@noop {} {\bibfield  {journal} {\bibinfo  {journal} {J. Phys. B: At.
  Mol. Opt. Phys.}\ }\textbf {\bibinfo {volume} {53}},\ \bibinfo {pages}
  {025301} (\bibinfo {year} {2020})}\BibitemShut {NoStop}%
\bibitem [{\citenamefont {Wright}\ \emph {et~al.}(2000)\citenamefont {Wright},
  \citenamefont {Arlt},\ and\ \citenamefont {Dholakia}}]{lag}%
  \BibitemOpen
  \bibfield  {author} {\bibinfo {author} {\bibfnamefont {E.~M.}\ \bibnamefont
  {Wright}}, \bibinfo {author} {\bibfnamefont {J.}~\bibnamefont {Arlt}},\ and\
  \bibinfo {author} {\bibfnamefont {K.}~\bibnamefont {Dholakia}},\ }\href@noop
  {} {\bibfield  {journal} {\bibinfo  {journal} {Phys. Rev. A}\ }\textbf
  {\bibinfo {volume} {63}},\ \bibinfo {pages} {013608} (\bibinfo {year}
  {2000})}\BibitemShut {NoStop}%
\bibitem [{\citenamefont {Castin}\ and\ \citenamefont {Dum}(1999)}]{castin}%
  \BibitemOpen
  \bibfield  {author} {\bibinfo {author} {\bibfnamefont {Y.}~\bibnamefont
  {Castin}}\ and\ \bibinfo {author} {\bibfnamefont {R.}~\bibnamefont {Dum}},\
  }\href@noop {} {\bibfield  {journal} {\bibinfo  {journal} {Eur. Phys. J. D}\
  }\textbf {\bibinfo {volume} {7}},\ \bibinfo {pages} {399} (\bibinfo {year}
  {1999})}\BibitemShut {NoStop}%
\bibitem [{\citenamefont {Kumar}\ \emph {et~al.}(2019)\citenamefont {Kumar},
  \citenamefont {{\relax Lon\v{c}ar}}, \citenamefont {Muruganandam},
  \citenamefont {Adhikari},\ and\ \citenamefont {{\relax
  Bala\v{z}}}}]{kumar19}%
  \BibitemOpen
  \bibfield  {author} {\bibinfo {author} {\bibfnamefont {R.~K.}\ \bibnamefont
  {Kumar}}, \bibinfo {author} {\bibfnamefont {V.}~\bibnamefont {{\relax
  Lon\v{c}ar}}}, \bibinfo {author} {\bibfnamefont {P.}~\bibnamefont
  {Muruganandam}}, \bibinfo {author} {\bibfnamefont {S.~K.}\ \bibnamefont
  {Adhikari}},\ and\ \bibinfo {author} {\bibfnamefont {A.}~\bibnamefont
  {{\relax Bala\v{z}}}},\ }\href@noop {} {\bibfield  {journal} {\bibinfo
  {journal} {Comput. Phys. Comm.}\ }\textbf {\bibinfo {volume} {240}},\
  \bibinfo {pages} {74} (\bibinfo {year} {2019})}\BibitemShut {NoStop}%
\bibitem [{\citenamefont {Dutta}\ \emph {et~al.}(2015)\citenamefont {Dutta},
  \citenamefont {Gajda}, \citenamefont {Hauke}, \citenamefont {Lewenstein},
  \citenamefont {{\relax D.-S. L\"uhmann}}, \citenamefont {Malomed},
  \citenamefont {{\relax T. Sowi\'nski}},\ and\ \citenamefont
  {Zakrzewski}}]{duttar}%
  \BibitemOpen
  \bibfield  {author} {\bibinfo {author} {\bibfnamefont {O.}~\bibnamefont
  {Dutta}}, \bibinfo {author} {\bibfnamefont {M.}~\bibnamefont {Gajda}},
  \bibinfo {author} {\bibfnamefont {P.}~\bibnamefont {Hauke}}, \bibinfo
  {author} {\bibfnamefont {M.}~\bibnamefont {Lewenstein}}, \bibinfo {author}
  {\bibnamefont {{\relax D.-S. L\"uhmann}}}, \bibinfo {author} {\bibfnamefont
  {B.~A.}\ \bibnamefont {Malomed}}, \bibinfo {author} {\bibnamefont {{\relax T.
  Sowi\'nski}}},\ and\ \bibinfo {author} {\bibfnamefont {J.}~\bibnamefont
  {Zakrzewski}},\ }\href@noop {} {\bibfield  {journal} {\bibinfo  {journal}
  {Rep. Prog. Phys.}\ }\textbf {\bibinfo {volume} {78}},\ \bibinfo {pages}
  {066001} (\bibinfo {year} {2015})}\BibitemShut {NoStop}%
\end{thebibliography}
\providecommand{\noopsort}[1]{}\providecommand{\singleletter}[1]{#1}%

\end{document}